%% file: paper-v1.tex
\documentclass[10pt,journal,compsoc]{IEEEtran}

\usepackage[pdftex]{graphicx}
\usepackage{url}
\usepackage{pdfcomment}
\usepackage{color}

\ifCLASSOPTIONcompsoc
  \usepackage[caption=false,font=footnotesize,labelfont=sf,textfont=sf]{subfig}
\else
  \usepackage[caption=false,font=footnotesize]{subfig}
\fi

\usepackage[linesnumbered,ruled]{algorithm2e}
\usepackage{multirow}
\let\oldnl\nl% Store \nl in \oldnl
\newcommand{\nonl}{\renewcommand{\nl}{\let\nl\oldnl}}% Remove line number for one line
\usepackage{algpseudocode}
\algnewcommand{\AND}{\textbf{ and }}
\algnewcommand{\Or}{\textbf{ or }}

\usepackage{amsmath}
\usepackage{enumitem}

\usepackage{tikz}
\usetikzlibrary{shapes}

\begin{document}

\title{DYVERSE: \textbf{DY}namic \textbf{VER}tical \textbf{S}caling\\in Multi-tenant \textbf{E}dge Environments}

\author{Nan~Wang,
       Michail~Matthaiou,
       Dimitrios S. Nikolopoulos~and~
       Blesson Varghese

\IEEEcompsocitemizethanks{\IEEEcompsocthanksitem The authors are with Queen's University Belfast, UK.\protect\\
E-mail: \{nwang03, m.matthaiou, d.nikolopoulos, b.varghese\}@qub.ac.uk
}
\thanks{Manuscript received September 17, 2018; revised month date, year.}}

%\markboth{IEEE Transactions on Services Computing,~Vol.~X, No.~Y, September~2018}%
%{Wang \MakeLowercase{\textit{et al.}}: DYVERSE: DYnamic VERtical Scaling in Multi-tenant Edge Environments}


\include{sections/abstract}

\maketitle
%\IEEEdisplaynontitleabstractindextext

%\IEEEraisesectionheading{
\section{Introduction}
\label{sec:introduction}
\input{sections/introduction}

\section{Background and Problem Model}
\label{sec:background}
\input{sections/background}

\section{Priority Management}
\label{sec:priority}
\input{sections/priority}

\section{Dynamic Vertical Scaling}
\label{sec:autoscaling}
\input{sections/autoscaling}

\section{Experimental Evaluation}
\label{sec:evaluation}

\input{sections/evaluation}

\section{Related Work}
\label{sec:relatedwork}
\input{sections/relatedwork}

\section{Conclusions}
\label{sec:conclusions}
\input{sections/conclusions}

%\section*{Acknowledgments}
%\input{sections/acknowledgments.tex}

\ifCLASSOPTIONcaptionsoff
  \newpage
\fi

%\begin{thebibliography}{1}


%\end{thebibliography}

\bibliographystyle{IEEEtran}  
\bibliography{references}

\end{document}

%% file: sections/abstract.tex
\IEEEtitleabstractindextext{%
\begin{abstract}
Multi-tenancy in resource-constrained environments is a key challenge in Edge computing. In this paper, we develop 'DYVERSE: DYnamic VERtical Scaling in Edge' environments, which is the first light-weight and dynamic vertical scaling mechanism for managing resources allocated to applications for facilitating multi-tenancy in Edge environments. To enable dynamic vertical scaling, one static and three dynamic priority management approaches that are workload-aware, community-aware and system-aware, respectively are proposed. This research advocates that dynamic vertical scaling and priority management approaches reduce Service Level Objective (SLO) violation rates.
An online-game and a face detection workload in a Cloud-Edge test-bed are used to validate the research. The merit of DYVERSE is that there is only a sub-second overhead per Edge server when 32 Edge servers are deployed on a single Edge node. When compared to executing applications on the Edge servers without dynamic vertical scaling, static priorities and dynamic priorities reduce SLO violation rates of requests by up to 4\% and 12\% for the online game, respectively, and in both cases 6\% for the face detection workload. Moreover, for both workloads, the system-aware dynamic vertical scaling method effectively reduces the latency of non-violated requests, when compared to other methods.
\end{abstract}

\begin{IEEEkeywords}
Edge computing, multi-tenancy, vertical scaling, dynamic priority, dynamic scaling
\end{IEEEkeywords}}

%% file: sections/introduction.tex
The vision of next-generation distributed computing is to harness the network edge for computing~\cite{edgecomputing-00, FeasibilityOfFog}. In contrast to servicing all user requests from the Cloud, a workload may be distributed across the Cloud and nodes, such as routers and switches or micro data centres, that are located at the edge of the network~\cite{edgecomputing-01, cardellini2019new, CloudFuturology}.
 
\begin{figure}
  \centering
  \includegraphics[width=0.5\textwidth]{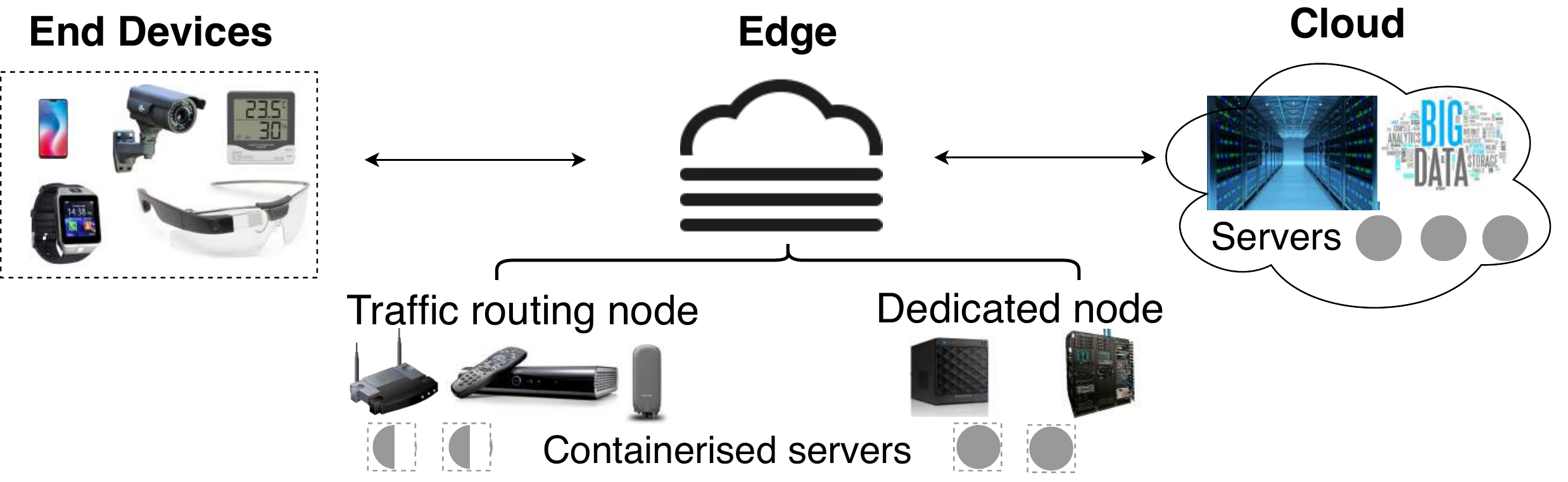}
  \caption{Example of a three-tier Edge computing architecture.}
  \label{fig:edgeComputing}
\end{figure}

Figure~\ref{fig:edgeComputing} shows a three-tier Edge computing architecture. The Cloud tier is represented by data centres that provide compute resources for workloads. The Edge tier uses nodes that are closer to users. These include: (i) traffic routing nodes -- existing nodes that route Internet traffic, for example, Wi-Fi routers, which may be augmented with additional compute resources, and (ii) dedicated nodes -- additional micro data centres, for example, cloudlets. Workloads hosted on the Edge could either be the same or a subset of functionalities of those hosted on the Cloud based on the availability of resources at the Edge. The end device tier represents user devices and sensors; 29 billion of these are estimated to be connected to the Internet by 2022\footnote{\url{https://www.ericsson.com/en/mobility-report/internet-of-things-forecast}}.

In Edge computing, end devices are connected to Edge nodes instead of directly to servers in the Cloud. The benefits of distributing a workload across the Cloud and the Edge have already been established in the literature. They include reduced communication latencies and reduced traffic to the Cloud, which in turn improves response times and Quality-of-Service (QoS)~\cite{taleb2017mobile, enorm}. 
There are challenges in achieving the vision of using Edge computing for distributing Cloud workloads, especially when Edge nodes are resource constrained as in traffic routing nodes. {\color{black}This paper aims to investigate the problem of supporting \textit{multi-tenancy} in Edge computing nodes with limited hardware resources when compared to the Cloud. In this case, the servers of multiple workloads hosted on the same Edge node are anticipated to compete for insufficient resources~\cite{taleb2017multi}. Similar to Cloud computing, multiple workloads are expected to share hardware resources in Edge computing~\cite{liu2016paradrop} since it would not be cost-effective to customise individual Edge computing nodes for a specific application~\cite{EdgeMarket}. Although when multiple workloads are asking for computing resources from a resource-constrained Edge node, a few of the workloads may continue using Cloud servers. This would minimise the opportunity to leverage the edge of the network for reducing the distance of data transfer and subsequently improving the QoS of applications~\cite{zhu2014improving, defog}. Therefore, it is important to efficiently support multi-tenancy on the Edge.}

Multi-tenancy {\color{black}on the Cloud is well researched} to host multiple Virtual Machines (VMs) on the same underlying hardware~\cite{liu2014reciprocal, bernabe2014semantic}. Nonetheless, it needs to be revisited in the context of Edge computing since: (i) Edge resources have limited processing capabilities, due to small form factor and low-power processors when compared to data centre resources~\cite{ahmed2016survey}. (ii)~The Edge is a more transient environment (availability of resources change over time and may be available only for short periods) compared to the Cloud~\cite{satyanarayanan2017emergence}. 

Multi-tenancy causes resources contention. 
Mechanisms employed on the Cloud to mitigate resource contention include \textit{vertical scaling}, which is a process of allocating/deallocating resources to/from workloads so that multiple workloads can coexist~\cite{verticalscaling-1}. 
These solutions are not suited for Edge environments since: (i) The mechanisms to monitor and optimise the allocation/deallocation of resources to meet user-defined objectives, specified as \textit{Service Level Objectives (SLOs)} are typically computationally intensive~\cite{slo-1, liu2018cloud}. (ii) Predictive models used for estimating resource demands will have insufficient data for training~\cite{Thoth2017, rahmanian2018learning}. Edge services are expected to have short life cycles and may result in insufficient data to feed into an accurate Machine-Learning (ML) model; (iii) Edge environments are expected to have more transient system states compared to the Cloud and Edge workloads may run in a come-and-go style. Solutions designed for the Cloud platforms do not consider this. Therefore, a light-weight vertical scaling mechanism to facilitate multi-tenancy on the Edge is required, which is proposed and presented in this article. 

%The availability of resources on Edge nodes will vary over time.
As the execution of a workload progresses in a multi-tenant environment, there may be workloads that require more or fewer resources to meet their SLOs. If required resources are not available for Edge workloads, then SLOs will be violated. A static resource provisioning method is unsuitable given the frequent changes in an Edge environment and resource scaling is therefore required. 
The research in this paper proposes a dynamic vertical scaling mechanism that can be employed in a multi-tenant and resource-limited Edge environment. 

The proposed mechanism is underpinned by a model that accounts for \textit{static priorities} (set before the execution of a workload) and \textit{dynamic priorities} (that changes during execution) of workloads on the Edge. While priorities have been exploited in Cloud computing~\cite{han2012lightweight, ccavdar2015simulation}, we investigate it in the context of Edge computing in this article. The Edge is expected to be a premium service for Cloud workloads, and therefore, selecting Edge service users becomes important. We propose three dynamic priorities that are workload-aware, community-aware and system-aware to this end. We hypothesise that dynamic vertical scaling along with priority management approaches will improve the QoS of multi-tenant workloads and reduce SLO violation rates. 

{\color{black}
This paper makes the following novel contributions:
\begin{itemize}
\item[i.] The development of a framework for supporting multi-tenancy in a resource-constrained Edge computing node. In this framework, the Edge service QoS maximisation problem in a three-tier environment is formulated by considering the SLO violation rate, server co-location, dynamic vertical scaling, and priorities of workloads. 

% The Edge service QoS maximisation problem in a three tier environment by considering the SLO violation rate, server co-location, dynamic vertical scaling, and priorities of workloads is formulated by proposing DYVERSE~-~\textbf{DY}namic \textbf{VER}tical \textbf{S}caling in \textbf{E}dge environments.

\item[ii.] The design of the dynamic priority approaches for managing Edge applications in a multi-tenant environment, which accounts for Edge-specific characteristics and different economic models. Currently, research using priority management (for example the decision for offloading or task queuing) focuses on the pre-deployment phase. DYVERSE, on the other hand, applies priority management after deployment.

%A static priority and three dynamic priority management approaches are proposed. These approaches account for Edge-specific characteristics in the context of economic models for a multi-tenant environment. Currently, research using priority management (for example the decision for offloading or task queuing) focuses on the pre-deployment phase. DYVERSE on the other hand applies priority management after deployment.

\item[iii.] The development of the lightweight dynamic vertical scaling mechanism that adjusts resource allocations for prioritised Edge applications after deployment. Existing resource management techniques are heavyweight (require significant processing) since they are computationally intensive~\cite{hong2018resource}. DYVERSE offers resource management suited for resource-constrained Edge nodes.

%A lightweight dynamic vertical scaling mechanism that integrates the static and dynamic priority mechanisms is developed. Existing resource management techniques are heavyweight (require significant processing) since they are computationally intensive. DYVERSE offers resource management suited for resource-constrained Edge nodes.

\item[iv.] The evaluation of DYVERSE on two different use-cases in a realistic experimental environment. Much of existing research on resource management for Edge computing is evaluated using simulators~\cite{mao2016dynamic, ku20175g, yu2016joint}. The benefits of applying DYVERSE in an Edge node is on the contrary presented in a test-bed with a location-based mobile game and a real-time face detection application.
\end{itemize}
}

The feasibility of the proposed priority management approaches and dynamic vertical scaling mechanism is validated using an online-game and a face detection workload in an Edge environment. {\color{black} These workloads are a natural fit for using the Edge since they are latency critical - the response time is affected by the distance between the user device and the server.}
The merit of DYVERSE is observed in that they only have a sub-second overhead per Edge server when 32 servers are deployed on an Edge node. Additionally, we observe that scaling using static and dynamic priorities reduces the SLO violation rates of user requests by up to 4\% and 12\% for the online game (6\% for the face detection workload) respectively, when compared to executing workloads on the Edge node without dynamic vertical scaling. 
Moreover, the proposed dynamic vertical scaling with the system-aware dynamic priority approach improves the latencies of requests that are not violated. The key result is that our initial hypothesis is confirmed.
%- dynamic vertical scaling along with priority management approaches improves the QoS of multi-tenant applications and reduces SLO violation rates.

The remainder of this paper is organised as follows. Section~\ref{sec:background} presents the background and develops the problem model.
Section~\ref{sec:priority} proposes one static and three dynamic priority management approaches for a multi-tenant Edge environment. 
Section~\ref{sec:autoscaling} presents a dynamic vertical scaling mechanism for Edge nodes. 
Section~\ref{sec:evaluation} experimentally evaluates DYVERSE against a catalogue of metrics - system overhead, SLO violation rate and latency. 
Section~\ref{sec:relatedwork} highlights the related work.
Section~\ref{sec:conclusions} concludes this paper.

%% file: sections/background.tex
The architecture considered in this paper is based on a three-tier model (as shown in Figure~\ref{fig:edgeComputing}) in which compute resources are located at the edge of the network closer to end devices~\cite{etsi2016mobile}.
In the Cloud tier, workloads are hosted on servers in a data centre. To enable the use of the Cloud in conjunction with the Edge nodes, we deploy a Cloud Manager along with each server. This manager is responsible for offloading a workload onto an Edge node. The Edge tier comprises an Edge node with an Edge Manager communicating with the Cloud Manager and managing the Edge resources. On-demand Edge servers are deployed to service requests from end devices. 
In the end device tier, multiple devices are connected to the Cloud servers. When an Edge server is deployed, the end devices connect to the Edge servers. 

\begin{table*}[t]
\centering
\caption{Notation used in the Cloud-Edge model}
\begin{tabular}{c  l  c} \hline
\textbf{Notation} & \textbf{Description} & \textbf{Source} \\ \hline
\(S\) & A set of Edge servers $s$, each hosted in a container on an Edge node.& \multirow{ 3}{*}{Edge Manager}\\ 
\(PS_s\) & Priority score of Edge server $s$; it is calculated statically $SPS_s$ or dynamically $(w/c/s)DPS_s$ \\
\(uR\) & One unit of CPU and memory &\\
\hline
\(U_s\) & A set of users $u$ to connect to $s$ & \multirow{ 4}{*}{Cloud Manager}\\ 
\(L_s\) & The service level objective (in terms of latency) of Edge server $s$ & \\
\(donation_s\) & flag for the willingness of donating resources from Edge server $s$ & \\
$dThr_s$ & Threshold of $L_s$ to allow the scaling down of Edge server $s$ & \\
\hline
%\(nL_{s,u}\) & Average network latency of user $u$ on Edge server $s$ & \multirow{ 9}{*}{Monitor}\\ 
%\(cL_{s,u}\) & Average computing latency of user $u$ on Edge server $s$ & \\ 
\(aL_s\) & Average latency of Edge server $s$ & \multirow{ 6}{*}{Monitor}\\ 
\(R_s\) & CPU and memory used by Edge server $s$ on the Edge node& \\
%\(rR\) & CPU and memory released by terminating servers on the Edge node & \\ 
\(VR_s\) & Service level objective violation rate of Edge server $s$ in the previous round of dynamic vertical scaling& \\ 
\(aR_s\) & Resources to be added to Edge server $s$ for scaling up& \\ 
\(FR\) & Free CPU and memory on the Edge node available for $S$ & \\
\hline
\(decision\) & Flag for 'scaleup' or 'scaledown' containers on an Edge node & Auto-scaler \\ \hline
\end{tabular}
\label{table:mathnotation}
\end{table*}

In this paper, we consider an Edge node to be a multi-tenant environment hosting multiple workloads. It would not be feasible to build bespoke Edge systems for individual workloads to improve their QoS. Therefore, workloads will need to share an Edge node. Unlike Cloud computing, where additional resources can be purchased when needed, Edge environments are resource limited and hence multi-tenancy is challenging.  

To support multi-tenancy on an Edge node and given that a suitable economic model is available for utilising the Edge, workloads need to be prioritised for using resources. If they are not prioritised, they have to equally share resources; this is inefficient since resource demands may vary and will be affected by the workloads executed. Prioritising workloads makes it possible to assign resources based on the demand of each workload and reduces any under-utilisation of resources. In this research, a priority management mechanism for an Edge node that (i) maintains a Priority Score (PS) for each running workload, and (ii) determines which workload should be first scaled is proposed. 

Resource provisioning has been considered in Edge environments~\cite{liu2016paradrop}.
However, existing research focuses on managing resources during the scheduling and deployment of workloads, referred to as pre-deployment management. In a multi-tenant Edge environment, where resources are limited and the availability of free resources keeps changing, it is important to efficiently manage resources periodically to not only avoid overloading the Edge node but also ensuring that the overall QoS of Edge workloads is satisfactory. Post-deployment resource management refers to managing resources after workloads are deployed~\cite{pahl2015containers}. In this paper, we consider post-deployment management along with dynamic priorities to address the resource allocation problem for the multi-tenant Edge environment.

\subsection{Problem Model}
The notation to represent an Edge environment is shown in Table~\ref{table:mathnotation}.
{\color{black}Let $S=\{s_1, s_2, \dots, s_n\}$ be a set of latency-sensitive servers hosted on an Edge node $e$. Each server $s \in S$ is a subset of functionalities of the Cloud server that is offloaded onto the Edge node. When deploying $s$ onto the Edge node, its Cloud Manager provides $U_s$, a set of users that are to be serviced by $s$, a latency objective $L_s$, its willingness to donate resources $donation_s$, and the threshold for permitting a scale-down action $dThr_s$. Then the multi-tenant Edge node is modelled as
\begin{align}
s_n=<U_{s_n}, L_{s_n}, donation_{s_n}, dThr_{s_n}>, \forall s_n \in S
\end{align}
}
Our research employs four components on an Edge node to facilitate multi-tenancy: 
(i) \textit{Edge Manager} maintains the Edge server registry, makes decisions on starting and terminates Edge servers;
(ii) \textit{Monitor} periodically monitors a number of metrics related to each Edge server $s$. These metrics include:
(a)~CPU and memory usage $R_s$, 
(b)~average latency $aL_s$ for all users $u \in U$,
(c)~workload intensity, for example, number of requests $Request_s$, and 
(d)~scaling frequency.
(iii) \textit{Auto-scaler} dynamically makes decisions to scale up or down an Edge server $s$ based on its performance;
and (iv) \textit{Edge Server} $s$ interacts with end devices. 

The research objective is to: \textit{develop the mechanisms required for sharing computing resources of an Edge node among multiple workloads while minimising the latency of each workload and maximising its performance.
The overall performance of workloads on the Edge node $e$ is measured by the SLO violation rate of all servers $S$:}
\begin{align}
VR_e = \frac{\sum_{s \in S} Request_s [aL_s > L_s]}{\sum_{s \in S} Request_s}
\end{align}\label{eq:violation_rate}The equation above may be affected by varying the resources allocated to an Edge server and the sequence of allocating/deallocating resources on Edge servers. We hypothesise that dynamic resource allocation along with priorities will reduce the SLO violation rate.

%% file: sections/priority.tex
Uniform allocation of resources to multiple tenants on an Edge node can occur at the same time. However, this is a static allocation technique and does not consider the specific needs of individual tenants. 
Customised allocations cannot always proceed concurrently since Edge nodes are resource constrained.
Therefore, the sequence of allocating resources for running servers on an Edge node needs to be considered. 
In this paper, it is assumed that Edge and Cloud service providers are different. Hence, priority management, scaling schemes, and resource management in the Cloud and Edge are decoupled.
Pricing models also affect the priority of an Edge server. For instance, it may be unfair to assign the same priority to two tenants with similar computational requirements -- when one tenant pays for a fixed period and the other pays for the resource. In this case, a different priority needs to be assigned to the tenants.

In this section, we propose Dynamic Priority Management (DPM) approaches when using different pricing models. DPM is compared against a Static Priority Management (SPM) approach, which is used as a baseline. We envision Edge computing to be a utility-based service relying on the \textit{pay-for-$X$} (pay for what you use) principle. The following three pricing models are considered: (i) Pay-For-Resources (PFR) -- the resource used is paid for, (ii) Pay-For-Period (PFP) -- the time for utilising resources is paid for, and (iii) Hybrid~\cite{al2013cloud} -- combines both PFR and PFP models. 

\subsection{Static Priority Management}
\label{sec:spm}
We define SPM as a baseline in the priority management of multiple tenants. PS is the importance of an Edge server (high value means high priority and the server is provided resources before others), which is calculated when the server is launched. The PS remains the same from deployment until termination of the server. This is comparable to a flat rate model in resource pricing. 
SPM is a realistic approach in contrast to an approach using a pre-defined PS provided by the Cloud Manager since the manager would give the highest PS to its Edge server. SPM, on the other hand, is affected by several factors that are measured by the Edge Manager and can differentiate the servers on an Edge node.

\begin{table*}[ht]
\centering
\caption{Factors affecting static priority management}
\begin{tabular}{c c l c} \hline
\textbf{Factor} & \textbf{Notation} & \textbf{Description} & \textbf{Source} \\ \hline
Premium Service & $P_s$ & Price paid for purchasing priority on Edge server $s$& Cloud Manager \\ 
\hline
First Come First Serve & $ID_s$ & The ID of Edge server $s$, set by the sequence of the request for launching $s$ & \multirow{ 3}{*}{Edge Manager}\\ 
Ageing & $Age_s$ & No. of times Edge server $s$ has been rejected by an Edge node & \\
Loyalty & $Loyalty_s$ & No. of times Edge server $s$ has used services on an Edge node & \\
\hline
\end{tabular}
\label{table:staticPriority}
\end{table*}

Table~\ref{table:staticPriority} lists four factors affecting the Static Priority Score (SPS) of an Edge server. The Cloud Manager of each workload is allowed to purchase the premium service to gain a higher PS. This factor is measured by the price a Cloud Manager has paid for premium service $P_s$. {\color{black}Such purchased priority has been adopted in designing scheduling algorithms in Cloud computing~\cite{chen2013user}, which aims to guarantee
that users who pay more can experience higher quality service.} The Edge Manager maintains a record of the sequence in which requests were made for using Edge service; each Edge server is marked with an ordinal number $ID_s$ when the Cloud Manager requests an Edge service. {\color{black}This is in line with the first-come-first-serve policy widely adopted in Cloud resource provisioning~\cite{he2012elastic}.} However, using only $P_{s}$ and $ID_{s}$ may result in priorities becoming fixed and may deprive a workload of resources if it did not initially purchase a premium service. To avoid this, the Edge Manager also considers ageing $Age_s$, which is the number of times an Edge server has been denied resource access. {\color{black}Aging is implemented in many scheduling algorithms for Cloud computing as a technique to ensure that all jobs are eventually scheduled~\cite{pastorelli2015hfsp}.} The last factor considered by the Edge Manager is loyalty $Loyalty_s$, which is the number of times the Edge service was used. {\color{black}Loyalty is used as an incentive for Cloud service users to adopt the Edge computing model~\cite{gao2016fogroute}.}
$SPS$ for an Edge server $s$ is calculated as:\vspace{-5pt}
\begin{align}
SPS_s & = W_P * P_s + W_{ID} * \frac{1}{ID_s} + W_{Age} * Age_s \nonumber \\
& \qquad + W_{Loyalty} * Loyalty_s
\label{eq:sps}
\end{align}
\noindent where $W$ is the weight assigned to a factor. {\color{black}In this paper, $SPS_s$ is used as a baseline combining the factors that are considered to be important in the literature as indicated above on resource management. This makes it a better baseline than using a single factor to decide the priority of an Edge server.} 

{\color{black}SPM is advantageous in that in SPM the priority score of an Edge application is calculated only once when it is deployed. This solution entails low overhead and is straightforward to implement. It is also proved that static factors are useful for designing effective resource management solutions in the context of Cloud computing~\cite{chen2013user, he2012elastic, pastorelli2015hfsp, gao2016fogroute}. However,} a disadvantage of SPM is that the PS of an Edge server may not remain the same in the real world. Dynamic factors affecting the performance of Edge servers, for example, whether a resource-intensive workload is executed needs to be considered. Hence, dynamic priorities are proposed {\color{black}in this paper, the aim is to further improve the effectiveness of priority management required for Edge computing.}

\subsection{Dynamic Priority Management}
\label{sec:dpm}
Table~\ref{table:dynamicPriority} lists factors affecting the Dynamic Priority Score (DPS) of an Edge server. Since vertical scaling occurs periodically (presented in Section~\ref{sec:autoscaling}), the values of these factors are updated in every round of vertical scaling. This ensures that varying priorities are taken into account for scaling decisions. In DPM, three approaches are introduced. The first is a workload-aware approach in which the workload of an Edge server in the previous round of scaling will affect its priority in the next round. The Monitor records three workload related metrics of each Edge server $s$, which includes the number of requests $Request_s$ and users $|U_s|$ serviced, and the amount of data transferred between the user and the Edge server $Data_s$. 

\begin{table*}[t]
\centering
\caption{Factors affecting dynamic priority management}
\begin{tabular}{c c l c} \hline
\textbf{Factor} & \textbf{Notation} & \textbf{Description} & \textbf{Source} \\ \hline
\multirow{ 3}{*}{Workload} & $Reqest_s$ & No. of requests serviced by Edge server $s$ & \multirow{ 5}{*}{Monitor} \\
 & $|U_s|$ & No. of users serviced by Edge server $s$ & \\ 
 & $Data_s$ & Amount of data processed by Edge server $s$ &  \\  
%\hline
Reward & $Reward_s$ & No. of times Edge server $s$ has donated resources & \\ 
Penalty & $Scale_s$ & No. of times Edge server $s$ has been scaled on an Edge node \\ 
\hline
\end{tabular}
\label{table:dynamicPriority}
\end{table*}

The second DPM approach is community-aware in which a workload can donate resources to a shared resource pool. If an Edge server $s$ donates, then it is recorded by the Monitor in $Reward_s$. In the subsequent scaling round, the server that donated will be rewarded a higher PS. 

The third DPM approach is system-aware in which the adverse effect of continuously scaling on the Edge is mitigated. Frequent scaling may result in an unstable Edge node that is continuously aiming to meet the requirements of workloads, which cumulatively would result in large overheads. To avoid this the Monitor keeps a record of the number of times an Edge server $s$ has been scaled $Scale_s$. This is used to penalise the workload when calculating its PS in the next scaling round.

\subsubsection{Workload-aware Dynamic Priority Score (wDPS)}
\label{sec:wDPS}
Workload-related factors presented in Table~\ref{table:dynamicPriority} 
differentiate individual workloads on the Edge node. 
For different pricing models, the workload-aware DPS needs to be different to ensure fairness among Edge service users. For example, in the PFP pricing model, the priority of an Edge server should be reduced if it will require more resources for a given time. 
However, in the PFR and Hybrid pricing models, it is assumed that additional use of resources is already paid for. Therefore, workload-related factors are not considered for issuing penalties in these models.

In the PFR and Hybrid pricing models, wDPS of an Edge server $s$ is defined as: \begin{align}
wDPS_{s,PFR/Hybrid} & = SPS_s \nonumber	\\
     & \qquad {} + W_{Request} * Request_s + W_U * |U_s| \nonumber \\
     & \qquad + W_{Data} * Data_s
     \label{eq:wdps1}
\end{align}{\color{black}With the above Equation, $wDPS$ may deprive small workloads by assigning them lower priority scores and assigning large workloads higher priority scores. This could result in the PFR and Hybrid pricing models being an unfair system for applications with varying workloads. Hence, to address fairness, Equation~\ref{eq:wdps2} defines $wDPS$ in the PFP pricing model as:}
\begin{align}
wDPS_{s,PFP} & = SPS_s \nonumber	\\
     & \qquad {} + \frac{1}{W_{Request} * Request_s} + \frac{1}{W_U * |U_s|} \nonumber \\
     & \qquad + \frac{1}{W_{Data} * Data_s}
     \label{eq:wdps2}
\end{align}{\color{black}The above equation addresses fairness by assigning lower priority scores to large workloads. Through} implementing wDPS, the Edge Manager captures the impact of different workloads in Edge servers and adjusts vertical scaling (considered in Section~\ref{sec:autoscaling}). For example, for a large workload in PFR and Hybrid pricing models, its PS will be set higher in the next round of scaling since compared to other workloads it requires more resources to maintain good performance. However, in the PFP pricing model, this workload will receive lower priority in the next scaling round because more resources may have to be added to the workload without being paid for.

A limitation of wDPS is that workload-related factors are objective and not controlled by the Cloud Manager. Therefore, a subjective factor to volunteer resources in exchange for rewards is further proposed to enrich DPM.

\subsubsection{Community-aware Dynamic Priority Score (cDPS)}
\label{sec:cDPS}
%\textbf{\textit{Community-aware Dynamic Priority Score (cDPS)}}:
When an Edge server is scaled, it may be allocated the same resources as before, i.e. scaling was not required. This may occur when the Edge server has met its SLO but does not meet a specified threshold (considered in Section~\ref{sec:autoscaling}) to be scaled down. Under such circumstances, the Edge server could still be scaled down as long as the risk of degraded performance is acknowledged. Reward credit is given to an Edge server if in a previous scaling round it agreed to donate 
%one unit of resource 
to a shared resource pool given that this Edge server had met its SLO. This is derived from the community model in economics, which encourages voluntary contributions. In this paper, we implement resource donation as an incentive for Edge servers to share resources allocated to them on an Edge node. This is beneficial when no additional resources can be requested from the Edge node during busy hours since the resources may be adjusted locally within Edge services.

{\color{black}cDPS of an Edge server $s$ is defined as: \vspace{-5pt}
\begin{align}
cDPS_s & = wDPS_s + W_{Reward} * Reward_s
\label{eq:cdps}
\end{align}}With cDPS, Edge servers take partial control of their priority, which is in contrast to wDPS. Contrary to the premium service defined in SPM, the reward is free and could be added frequently in the life cycle of an Edge server. 

The above DPS' do not consider the impact of frequent scaling.
%, which is considered next
When many servers are running on an Edge node, continuous scaling could result in large overheads and a slower system response, which affects the performance of all Edge services. Hence, a penalty is imposed for frequently scaling in the next DPS approach.

\subsubsection{System-aware Dynamic Priority Score (sDPS)}
\label{sec:sDPS}
%\textbf{\textit{System-aware Dynamic Priority Score (sDPS)}}:
Frequent scaling may result in an increased overhead when consecutive scaling rounds occur between short time intervals. This is considered when updating the priority of all Edge servers. If an Edge server has to be scaled many times to not violate its SLO, then its PS will be set lower in the next scaling round as a penalty for slowing the system. One reason for frequent scaling may be due to unrealistic SLOs which are set by the Cloud Manager. Therefore, any adverse impact on the Edge node can be mitigated with a lower priority.
{\color{black}sDPS of an Edge server $s$ is defined as:\vspace{-6pt}
\begin{align}
sDPS_s & = cDPS_s + \frac{1}{W_{Scale} * Scale_s}
\label{eq:sdps}
\end{align}}By implementing sDPS, the Edge Manager penalises servers that slow down the Edge node.

%% file: sections/autoscaling.tex
Dynamic vertical scaling is a mechanism to allocate or deallocate Edge resources for a workload at runtime.
Efficient resource management is essential to better utilise Edge resources and ensure that the overall QoS is not compromised. Most resource management mechanisms consider resource provisioning during workload deployment but ignore the need for post-deployment resource adjustment (after the workload has started execution). 
Without an efficient dynamic vertical scaling technique, an Edge node could be overloaded when executing bursty workloads. 
If additional resources are not allocated to the workload, then SLO violations are likely to occur. Therefore, a mechanism that is constantly aware of resources on an Edge node and makes scaling decisions for resource allocations periodically is necessary for an Edge environment.

Since it is disadvantageous to scale resources for multiple workloads concurrently (as presented in Section~\ref{sec:priority}), a priority-based dynamic vertical scaling mechanism is required to reallocate resources for every Edge server. Workloads with the highest PS should be considered first and the one with the lowest PS at the end of a scaling round until there are no resources to support Edge servers with low priorities. Workloads that have insufficient resources will not be executed on the Edge node and will need to be executed elsewhere. 
In this research, workloads are deployed on the Edge node using LXC containers\footnote{\url{https://linuxcontainers.org/lxc/introduction/}}. 

\subsection{Priority-based dynamic vertical scaling}

Procedure~\ref{algo:autoscaling} presents a dynamic vertical scaling mechanism using priorities. The PS of a list of Edge servers executing on the Edge node is updated at the beginning of each round of vertical scaling (Line~1). The PS of an Edge server is calculated with the static and dynamic priority approaches defined in Section~\ref{sec:priority}. The list of servers is then sorted by the updated PS~(Line~2). The server with the highest priority is firstly considered by the Auto-scaler.
The Auto-scaler checks if there is a need for the Edge server on the node (whether users are covered by this Edge node or whether the QoS of the workload can be improved on the Edge node; Line~4). Network latency is used to decide whether the Edge server can deliver the desired improvement or whether terminating the Edge server and servicing the users through the original Cloud server can be of more benefit~(Line~20). In this paper, we assume that Edge nodes are not interconnected and therefore a workload cannot be migrated from one Edge node to another directly. Migration would need to occur via the Cloud. 
%The termination mechanism used here is the same as presented in Procedure~\ref{algo:termination}.

\begin{algorithm}
\SetAlgorithmName{Procedure}{}
\DontPrintSemicolon
 \KwData{\(S, L_s, uR, donation_s\)}
  Update $PS$ of all running Edge servers\;
 Sort $S$ by $PS$\;
 \For{$\forall s \in S$}{ 
   \eIf{ 
   $s$ is active \AND network latency is acceptable}{
   		\eIf{$aL_s > {L_s}$}{
            calculate $VR_s$\;
      		$scale(scaleup, VR_s, S)$\;
        	}{
           	\eIf{$aL_s > dThr*{L_s}$}{
        \eIf{$donation_s==1$}{
            	$scale(scaledown, uR, S)$\;
        }{	no scaling for $s$\;
        }}{
        	$scale(scaledown, uR, S)$\;
        }}}
   {
   		$terminate(S, s, U_s)$\;
   }}
\caption{Dynamic vertical scaling mechanism}
 \label{algo:autoscaling}
\end{algorithm}

The average latency is compared with the SLO provided by the Cloud Manager in its service request~(Line~5). If the latency is higher than the SLO (i.e. the Edge server has not been performing as expected), then the container hosting the Edge server will be allocated more resources. The ratio of resources to be added is based on the SLO violation rate of the server~(Line~6). This is to make sure that resources are adjusted based on a server's performance. If the latency is lower than a pre-defined percentage $dThr$ of the SLO, for example, 80\%, then resources are removed from the container hosting this server~(Line~16). When latency is between the threshold and the objective, resources are not scaled down since performance may be negatively affected~(Line~13). However, if the server is willing to donate resources in exchange for priority credits, it is scaled down and the number of times that the Edge server has donated resources is updated~(Lines~10-11). If resources of a server are scaled, then the total number of times it has scaled is recorded. However, note that the scaling that occurs when resources are donated is not recorded, since it is not used for penalising a server. {\color{black}Sorting by $PS$ in Line~2 of the above procedure has a computational complexity of $O(n\log{}n)$, where $n$ is the number of running servers on an Edge node. The priority-based dynamic vertical scaling loops through the $n$ running Edge servers, resulting in a complexity of $O(n)$. The total computation complexity of applying DYVERSE on an Edge node is then $O(n\log{}n)$ if there are 10 or more running Edge servers, or $O(n)$ if there are less than 10 running Edge servers; the actual complexity may be less since some of the low-priority Edge servers may be terminated before one round of the priority-based dynamic vertical scaling completes.}
%\todo{BV: Could you please check if the rephrase is correct}

%\textit{Multi-tenant Management}: 
\subsection{Multi-tenant Management}
When a server is scaled, then the scaling mechanism as shown in Procedure~\ref{algo:scaling} first checks the decision on whether to ``scaleup'' or ``scaledown''~(Lines~1 and 17). To scale up, i.e. to allocate more resources, the Monitor firstly calculates the amount of resources to add~(Line~3) and then checks if there are additional available resources on the Edge node to support this~(Line~4). If resources are available, then they are added to the container~(Line~5). Re-configuring the resource limits of a container is realised through the \(cgroup\) command, i.e. control group, which is a feature of Linux kernel that limits, accounts for and isolates the resource usage (for example, CPU and memory) of a container\footnote{\url{https://linuxcontainers.org/lxc/manpages/man1/lxc-cgroup.1.html}}. If the available resource is not sufficient to support scaling up, then the container with the lowest priority in the container set will be terminated to release its resources so that there are more resources available~(Line~8). During this process services for users connected to this Edge server are still available through the Cloud server. The containers with lower priorities will be terminated sequentially until the updated $FR$ is sufficient to support the scale-up request or there are no more containers with lower priorities~(Lines~10-13). To scale down, a unit of resource is removed from the server~(Line~18). At the end of the scaling process, the Edge server is updated with a new quota of resources~(Lines~5, 14 and 18).  
%When an Edge server is either scaled or migrated, the Auto-scaler takes the next server from the list to repeat the process above until all Edge servers are checked. 

\begin{algorithm}
\SetAlgorithmName{Procedure}{}
\DontPrintSemicolon
 \KwData{\(decision, uR/VR_s, S\)}
 \If{\( decision == scale up\)}{
 Measure $R_s$, $FR$\;
 $aR_s = R_s * VR_s$\;
  \eIf{\( FR >= aR_s\)}{
   Add $aR_s$ to $s$\;
   }{ 
   \While{\(FR < aR_s\)}{
   terminate($s_n, S, U_{s_n})$\; \tcp{\footnotesize$s_n$ is the last server in $S$}
   measure $FR$\;
   \If{\( n==\) index of s}{
   $ {\bf break} $\;}
 }
 Add $FR$ to $s$\;
 }  
   }
   \If{$decision == scale down$ }{
 Remove $uR$ from $s$\;
 }
\caption{Scaling mechanism}
 \label{algo:scaling}
\end{algorithm}

{\color{black}
\begin{figure}
  \centering
  \includegraphics[width=0.5\textwidth]{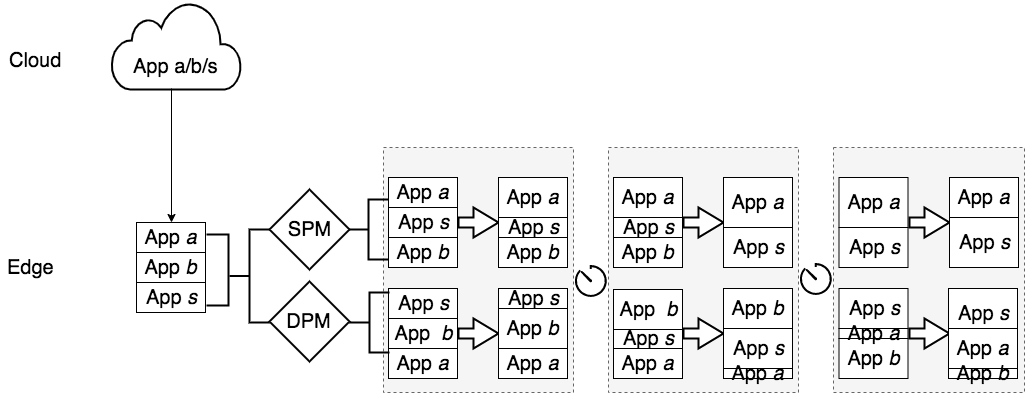}
  \caption{Example application of DYVERSE on an Edge node.}
  \label{fig:dyverse}
\end{figure}

Figure~\ref{fig:dyverse} illustrates the application of DYVERSE on an Edge node using an example. When applications (for example App $a$, $b$ and $s$) from the Cloud are deployed into an Edge node, they have an initial priority level,
such as purchased priorities (for example, application $a$ has a higher priority than $b$ and $b$ has a higher priority than $s$.
The dynamic vertical scaling mechanism (marked in grey block) that was proposed in this section is carried out periodically, for example, every 5 minutes on the Edge node. If the Edge node is configured with the static priority management (SPM) defined in Section~\ref{sec:spm}, the order of these applications being considered in the first round of the dynamic vertical scaling is calculated with Equation~\ref{eq:sps}. For example, the order of applications in Figure~\ref{fig:dyverse} with SPM is $a$, $s$ and $b$. Then in the subsequent rounds, the priority scores of these applications stay the same, because with SPM the priority scores of applications are only calculated during deployment.
On the contrary, if the Edge node is configured with the dynamic priority management (DPM) defined in Section~\ref{sec:dpm}, the order of these applications being considered in every round of the dynamic vertical scaling is updated. For instance, the order of applications in Figure~\ref{fig:dyverse} with DPM in the first round is $s$, $b$ and $a$. In the second round of dynamic vertical scaling, the order is updated to  $b$, $s$, and $a$. The dynamic priority scores of these applications are calculated using either Equation~\ref{eq:wdps1}, Equation~\ref{eq:wdps2}, Equation~\ref{eq:cdps} or Equation~\ref{eq:sdps}, depending on the selected DPM approach.

When Edge applications are initially deployed, an equal amount of resources are allocated to them,
which is represented by a uniform size of the box. 
Before each round of dynamic vertical scaling (grey block), the order of Edge applications are either kept the same by SPM or updated by DPM, then the resources allocated to each of the applications are checked by Procedure~\ref{algo:autoscaling}.
During the periodical dynamic vertical scaling, the resources allocated to Edge applications change over time, which is represented in Figure~\ref{fig:dyverse} with changing sizes of the box.
The decision to scale up/down or perform no scaling is based on monitoring the QoS of applications.
%\todo{BV: Please check if rephrase is right}. 
In addition, the selected priority management approach also has an impact on the dynamic vertical scaling.
For example, in the three rounds of dynamic vertical scaling illustrated in Figure~\ref{fig:dyverse}, application $b$ is completely removed from the Edge node when SPM is applied whereas it could have benefited from more Edge resources in the first two rounds when DPM is applied.}

%\textit{Discussion}:
\subsection{Discussion}
\label{termination}
In Edge computing, Edge servers when compared to Cloud servers will be used for shorter time intervals given the demand and limited availability of resources on Edge nodes. 
Therefore, in addition to dynamic vertical scaling, the Edge Manager will need to decide when an Edge server is removed from the Edge node.
%as shown in Procedure~\ref{algo:termination}.
The Edge Manager terminates (i.e., the service does not run on the Edge, but continues to be hosted on the Cloud server) service in the following three cases, which are considered in dynamic vertical scaling. Firstly, there are no free resources to support the Edge service. Secondly, the Edge service is not required anymore (the Edge server has been idle for a while). Thirdly, the Edge service does not improve the QoS of the workload (the performance constraints cannot be satisfied by an Edge server). 

When an Edge server is terminated, the associated data containing local updates are migrated to the Cloud (local data is appended to global data maintained by the Cloud Manager.
%; Line~1). 
This is realised through a key-value based data store, Redis\footnote{\url{https://redis.io/}}. {\color{black}Such a migration is use case dependent and only required by those applications that maintain data states across different geographic locations. For example, the Edge server of an application that processes video streams and does not require data to be stored could be terminated straightforwardly without requiring any data migration. In addition, for applications that require data migration,} we assume the benefit of using the Edge outweighs the communication costs since only user data is moved from the Edge to the Cloud, rather than the entire server. Servers that do not rely on databases would have minimal data movement. User devices are redirected to the Cloud server during the termination process. {\color{black} For example, in the case of a location-based mobile application, location-specific data is maintained on the Edge node and would need to be migrated to the Cloud when the application no longer executes on the Edge. It is noted that DYVERSE can benefit applications with different workflows in improving their overall QoS - applications that maintain their data state on the Edge and those in which the data is stateless on the Edge.}

%% file: sections/evaluation.tex
In this section, the priority management approaches presented in Section~\ref{sec:priority} and the dynamic vertical scaling mechanism proposed in Section~\ref{sec:autoscaling} are evaluated. The experimental setup, including the hardware platform and the distributed workloads employed in this research are firstly presented followed by the merits of DYVERSE
%the proposed priority approaches and scaling mechanism 
against metrics including system overhead, SLO violation rate, and latency.

\subsection{Setup}
A Cloud-Edge platform is set up using Amazon Web Services Elastic Compute Cloud (AWS EC2) and an ODROID-XU board. On the Cloud, t2.micro Virtual Machine (VM) running 14.04 LTS from the Dublin data centre is used to host the Cloud server. Although we employ a low-cost and basic VM, we note that in this paper we make no comparisons about the compute capabilities between the Cloud server and the Edge server. Therefore, in no way would the results presented in this paper be affected had we chosen a larger VM from the same data centre.
%The Cloud server maintains the global view of all the users. 

The Edge node is located in the Computer Science Building of Queen's University Belfast in Northern Ireland. The board has 2~GB of DRAM memory, and one ARM Big.LITTLE architecture Exynos 5 Octa processor running Ubuntu 14.04 LTS. Each server on the Edge node is hosted in an LXC container. %\color{black}When computing priority scores, the weights in Equation~\ref{eq:sps}, Equation~\ref{eq:wdps1}, Equation~\ref{eq:wdps2}, Equation~\ref{eq:cdps}, and Equation~\ref{eq:sdps} are set equal to 1.}

%that maintains local views of users connected to the Edge node. The global view on the Cloud server is updated periodically, but less frequently. The frequent location specific changes are updated locally. 

\subsection{Workloads}
%\textit{\textbf{Workloads: }}
Two workloads are chosen to evaluate DYVERSE: a location-based mobile game and a real-time face detection workload. Both workloads are server-based and a natural fit for Edge computing since they are latency critical~-- response time is affected by the distance between user devices and the server. Hence, a subset of the functionalities of the Cloud server can be brought closer to devices.

The above workloads are also representative of different workloads that can benefit from DYVERSE: the mobile game represents a multi-user application whose Edge server responds to incoming user requests; the face detection workload, in contrast, is representative of a data-intensive streaming application, in which case the Edge server pre-processes incoming data and relays it to the Cloud.
%use-case is employed as the workload in this paper for evaluating the proposed priority (both static and dynamic) management approaches and dynamic vertical scaling technique on Edge nodes. 

\textit{(i) Location-based mobile game}:
The application is an open-source 
%implementation of a 
game similar to Pok\'emon Go, named iPokeMon\footnote{\url{https://github.com/Kjuly/iPokeMon}}. iPokeMon comprises a client for the iOS platform, which can be used on mobile devices, and a server that is hosted on a public Cloud. 
User navigates through an environment in which virtual creatures named Pok\'emons are distributed. 
%It distributes virtual creatures named Pok\'emons on to the real world in which users navigate through an environment. 
The iPokeMon game server was redesigned to be hosted on the Cloud and an Edge node. The Edge hosts a latency-sensitive component that updates the virtual environment as a user navigates; for example, the GPS coordinates of the player and the Pok\'emons. The local view on the Edge server is updated by frequent requests sent to the Edge server. If user requests are serviced from a distant data centre, then user experience is affected due to lags in refreshing. Hence, the Edge is beneficial to reduce latencies for this workload. 
Up to a maximum of 32 Edge servers are hosted, with each server randomly supporting between 1 and 100 users.

The server on the Edge node is tested using Apache JMeter\footnote{\url{http://jmeter.apache.org/}}. One session of a connection (a user is playing the iPokeMon game) between the user device and the Edge server hosted in the LXC container is recorded for 20 minutes. During this time the number and type of requests and the parameters sent through the requests are recorded. Subsequently, JMeter load tests single and multiple Edge servers by creating virtual users and sending requests to the Edge servers from the virtual users in the experiments.

\textit{(ii) Real-time face detection}:
The original Cloud-based Face Detection (FD) workload captures video via a camera (such as CCTV footage) and transmits it to the Cloud where it is converted to greyscale (this is one-third the size of the color video) and then detects faces on it using OpenCV\footnote{\url{https://opencv.org/}}. Pre-processing (converting to greyscale) is performed on the Edge and hosted via LXC containers. The converted stream is sent to the Cloud for reducing the bandwidth used instead of sending colour video to the Cloud. Up to 32 servers are load tested on the Edge node, with each server randomly pre-processing between 0.1 and 1 frame per second.

The experimental evaluation demonstrates:
{\color{black}(i) the impact of varying priority factor weights on the priority score, }(ii) the overheads in priority management and the overheads of dynamic vertical scaling in a multi-tenant environment, (iii) the effect of DYVERSE on SLO violation rates, and (iv) the effect of DYVERSE on the latency of the above workloads.

\subsection{Results}
The experiments provide insight into the benefits of using the priority approaches and dynamic vertical scaling mechanism in a variety of scenarios, such as
%single and multiple users, 
moderate bandwidth-consuming and bandwidth-hungry tasks, single and multi-tenant servers and varying user-defined SLOs.  

\subsubsection{Impact of Priority Factor Weights}
\label{sec:weights}
In order to understand the impact of the static and dynamic factor weights introduced in Section~\ref{sec:priority}, we firstly study the priority scores of a particular Edge server when the weights are ranged in $[0, 1]$. The nine individual factor weights are grouped into four based on the priority management approach they belong to, namely the factor weights for $SPS$ in Section~\ref{sec:spm}, the factor weights for $wDPS$ in Section~\ref{sec:wDPS}, the factor weights for $cDPS$ in Section~\ref{sec:cDPS}, and the factor weights for $sDPS$ in Section~\ref{sec:sDPS}.

\begin{figure*}
\begin{center}
	\subfloat[]
	{\label{fig:heat_SPS}
	\includegraphics[width=0.3\textwidth]
	{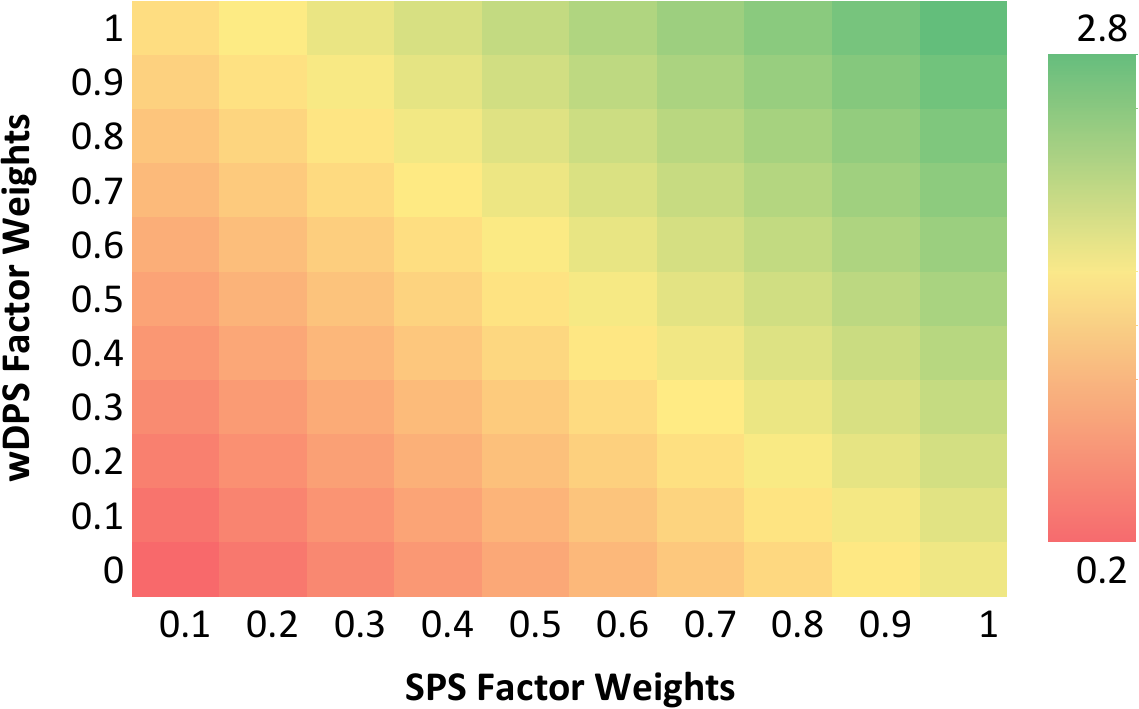}}
\hfill
	\subfloat[]
	{\label{fig:heat_wDPS}
	\includegraphics[width=0.3\textwidth]
	{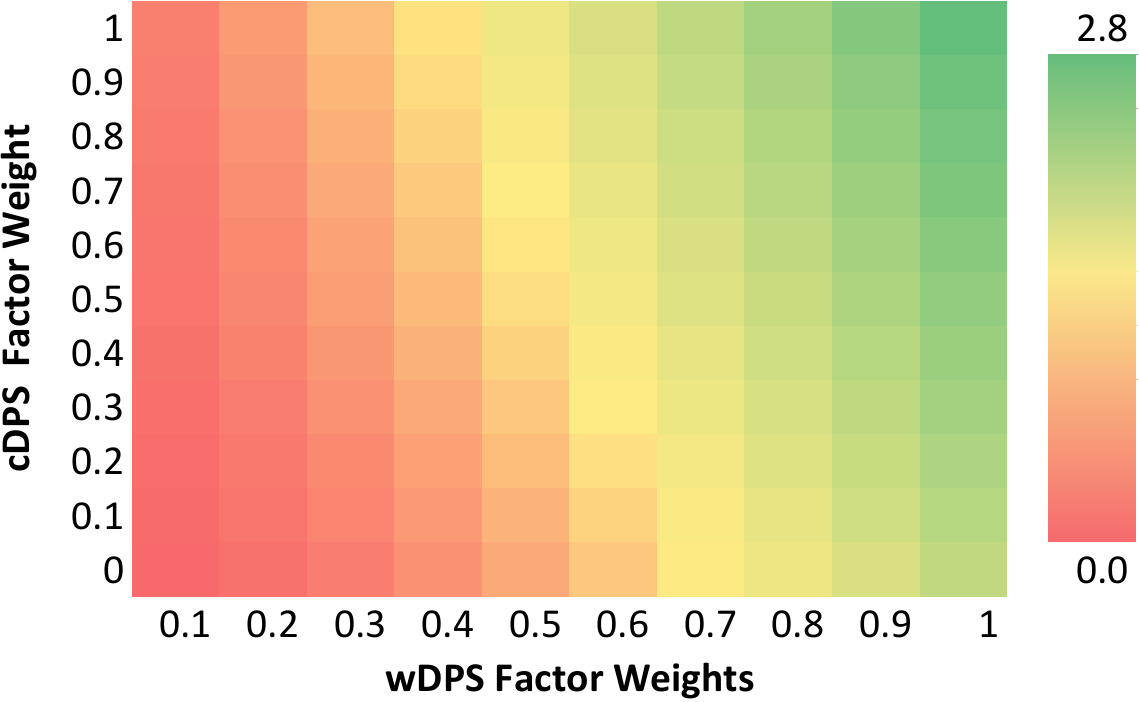}}
\hfill
	\subfloat[]
	{\label{fig:heat_cDPS}
	\includegraphics[width=0.3\textwidth]
	{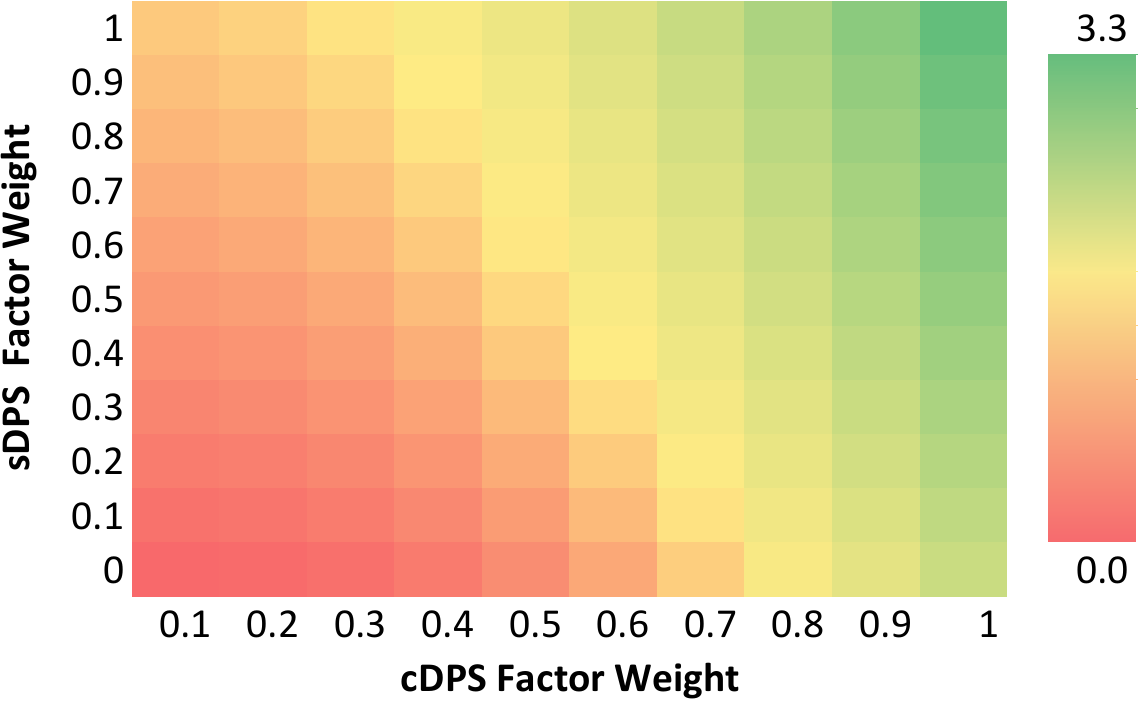}}
\end{center}
\caption{Priority scores of an Edge server when the priority factor weights vary.}
\label{fig:heat}
\end{figure*}

Figure~\ref{fig:heat} presents the priority scores of an example Edge server when the priority weights have different values. The x axis range in $[0.1, 1]$ because the dynamic priority approaches are developed incrementally from the static priority as shown in Equation~\ref{eq:sps}, Equation~\ref{eq:wdps1}, Equation~\ref{eq:wdps2}, Equation~\ref{eq:cdps} and Equation~\ref{eq:sdps}. The comparisons between each pair of directly related priority approaches show that when the factor weights increase, the priority of the Edge server increases. By comparing the three sub-figures, it is inferred that the influence of the factor weights on the priority score, from the strongest to the weakest, is $SPS$, $wDPS$, $sDPS$, and $cDPS$.

{\color{black}There is an indirect impact of different priority scores resulted from the different configurations of factor weights on the overall performance of an Edge system.}
{\color{black} %The impact of priority scores as a result of different configurations of factor weights could be indirectly passed\todo{BV: Please check this sentence - The impact of priority scores .... could be directly passed on to the overall performance?} on to the overall performance of an Edge system.
For example, in Figure~\ref{fig:heat_SPS} when the $wDPS$ factor weight is set as 0, the priority score is the same as adopting the $SPS$ approach.}
{\color{black}When the $wDPS$ factor weight is set lower than the $SPS$ factor weight, the priority score is mainly affected by the four factors defined in Table~\ref{table:staticPriority}, whereas the workload factors defined in Table~\ref{table:dynamicPriority} have less impact on the priority score. This could lead to a situation when an idle Edge application has a similar priority score as a busy Edge application that is servicing a large number of user requests, which further causes unfairness.}
{\color{black}%When the $wDPS$ factor weight is set lower than the $SPS$ factor weight, the priority score is dominant\todo{BV: Rephrase - the priority score is dominant. Do you mean the priority score is only affected by the four factors...?} by the four factors defined in Table~\ref{table:staticPriority} and the workload factors defined in Table~\ref{table:dynamicPriority} have less impact on the priority score. This could lead to a situation when an idle Edge application has a similar priority score as an overloaded Edge application\todo{BV: not sure what overloaded Edge application is?}, which further causes unfairness.

DYVERSE aims to reduce the average SLO violate rate for all running applications on an Edge node, as defined in Equation~\ref{eq:violation_rate}. Therefore, it is necessary to consider both application-dependent factors (e.g. workload factors) and system-related factors (e.g. reward and penalty.} In the following experiments, the weights of Equation~\ref{eq:sps} to Equation~\ref{eq:sdps} for calculating the priority scores are set to 1 to balance the impact of the different groups of factors.

\subsubsection{System Overhead}

% \begin{figure*}
% \begin{center}
% 	\subfloat[Mixed user behavior]
% 	{\label{fig:figure2a}
% 	\includegraphics[width=0.49\textwidth]
% 	{images/figure2a.pdf}}
% \hfill
% 	\subfloat[Aggressive user behavior]
% 	{\label{fig:figure2b}
% 	\includegraphics[width=0.49\textwidth]
% 	{images/figure2b.pdf}}
% \end{center}
% \caption{Overhead of dynamic vertical scaling per Edge server for varying user behavior.}
% \label{fig:figure2}
% \end{figure*}

\begin{figure}
\begin{center}
	\subfloat[Priority management]
	{\label{fig:figure1a}
	\includegraphics[width=0.5\textwidth]
	{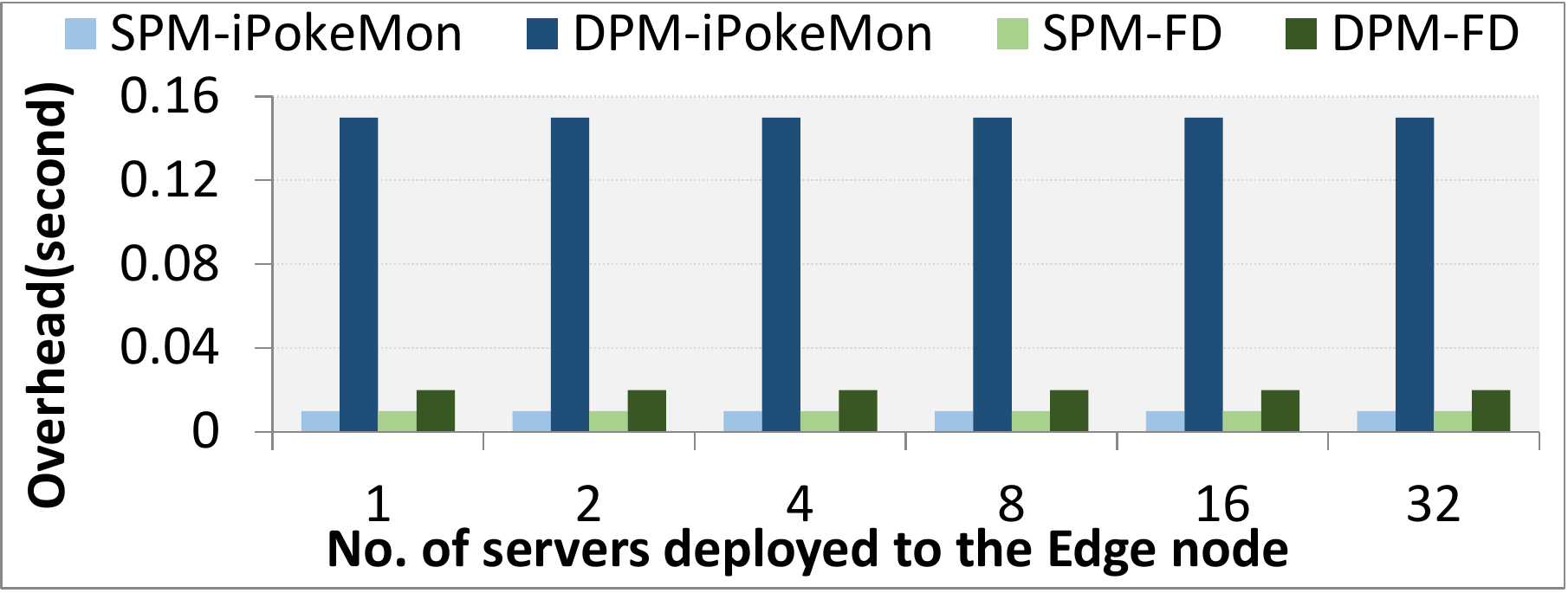}}
\hfill
	\subfloat[Dynamic vertical scaling]
	{\label{fig:figure1b}
	\includegraphics[width=0.5\textwidth]
	{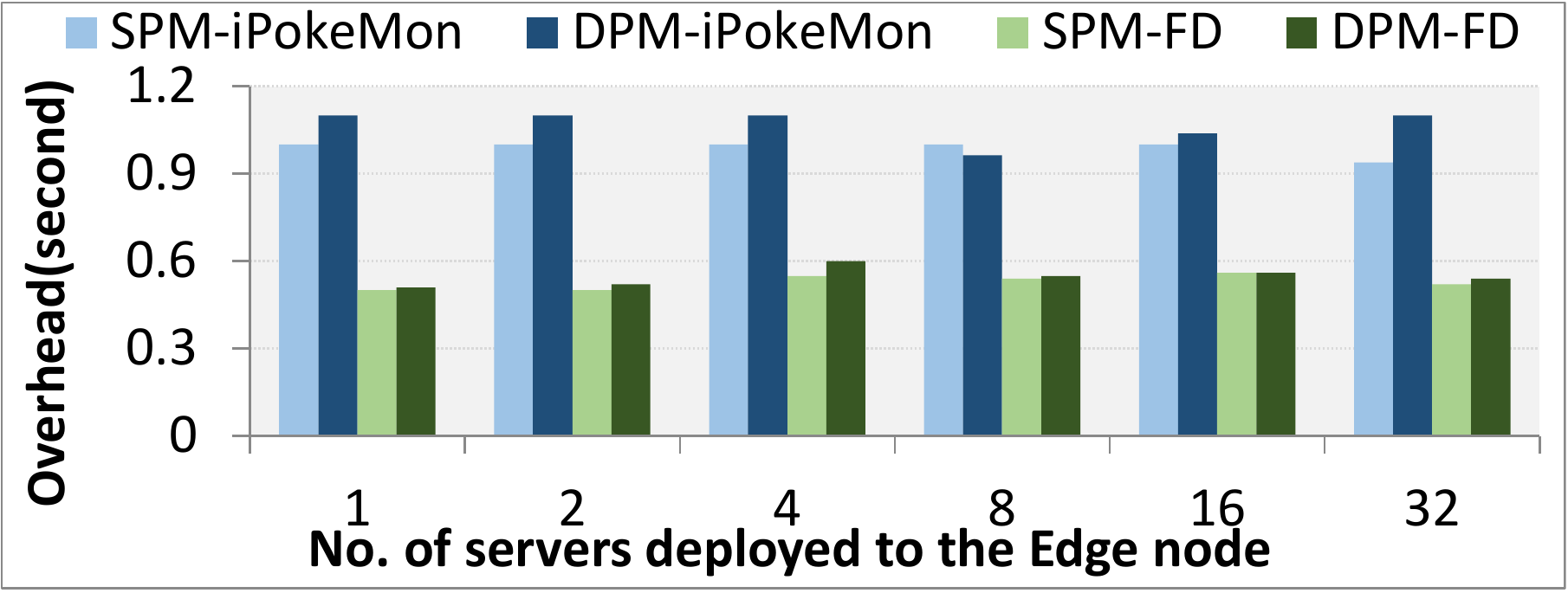}}
\end{center}
\caption{Overhead of priority management and dynamic vertical scaling per Edge server for iPokeMon and Face Detection (FD).}
\label{fig:figure1}
\end{figure}

Edge platforms are expected to have more transient system states than the Cloud, i.e. Edge workloads have shorter life cycles, so lower overheads in resource management are better. We measured the overheads of priority management (Figure~\ref{fig:figure1a}) and dynamic vertical scaling (Figure~\ref{fig:figure1b}) of each server on the Edge node for iPokeMon and FD. It should be noted that the servers are still servicing requests 
%(and are not decommissioned) 
when priorities are updated. Similarly, the overhead of dynamic vertical scaling, which is the time taken to reallocate the resources of the server was measured. During this time the servers continue to service requests. Although three dynamic priorities (wDPS, cDPS, and sDPS) were proposed, only sDPS is considered for comparing the overheads between SPM and DPM; we noted that different dynamic priorities did not affect the overall delay in the experiments. 

Figure~\ref{fig:figure1a} shows that SPM incurs small overheads, but it is more costly to employ DPM, especially for iPokeMon. For example, for 32 servers the overhead of using DPM with iPokeMon is 150~milliseconds (ms) when compared to 10ms for SPM. This is because static priorities assigned to servers do not change during execution and monitoring is needed for dynamic priorities.
%system- and workload-related 
%additional factors have to be monitored. 
iPokeMon server records all requests and responses in a log file while supporting multiple users and the Monitor reads through this file in every scaling round to update the dynamic priority. The FD server maintains the values of dynamic factors while processing requests from the single streaming source and the Monitor uses these values to calculate the PS. One insight from the difference in DPM overheads is that monitoring workloads could be accounted for when designing Edge applications.

%The average overhead for each server decreases with increasing number of servers in aggressive user behavior. This is because a number of servers with lower priorities will be terminated on the Edge node. This happens when there are no free resources available on the Edge node; higher priority servers will be allocated resources that are released from servers with lower priorities. So, if the execution started with 16 servers, then at the end of vertical scaling, there may be fewer servers than when execution started. This reduces the average overhead. 

The overhead of dynamic vertical scaling in Figure~\ref{fig:figure1b} using DPM is on average higher than using SPM. This is because, during scaling, dynamic factors such as reward and penalty are monitored for each server. {\color{black}More time is required to scale iPokeMon servers because during termination user data needs to be migrated from the Edge to the Cloud. When an FD server is terminated no data is migrated. As described in Section~\ref{termination}, the termination procedure is use case dependent and only applies to iPokeMon in this paper. The application of DYVERSE on two use cases is to demonstrate the different types of workflows that can benefit from dynamic vertical scaling post-deployment of the application.}

\begin{figure}
  \centering
  \includegraphics[width=0.5\textwidth]{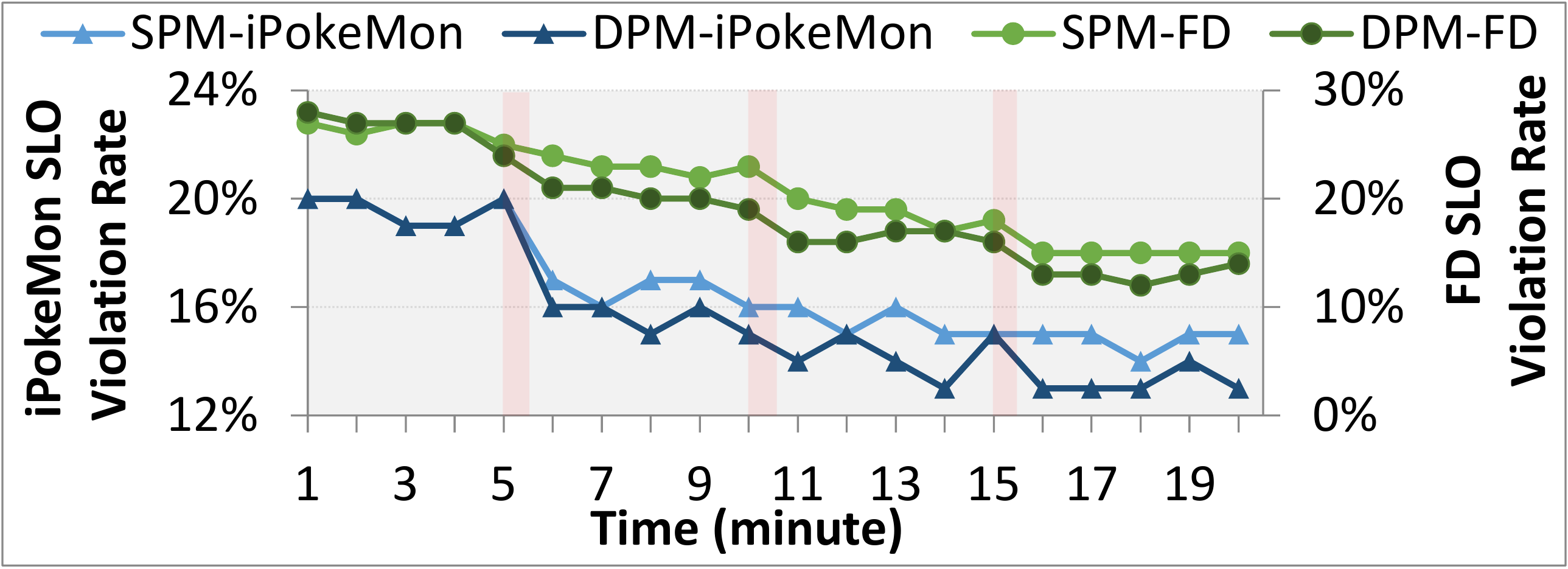}
  \caption{\color{black}Violation rate when the SLO is 78ms for iPokeMon and 2.13s for FD. 32 Edge servers are deployed. Red blocks correspond to the overhead time when priority-based scaling occurs.}
  \label{fig:overheadSLO}
\end{figure}

To understand the impact of the observed overheads on the QoS of iPokeMon and FD, the average SLO violation rate of 32 Edge servers every one minute is presented in Figure~\ref{fig:overheadSLO}. The first five minutes window is the period when the Edge servers are launched with an equal amount of resources. During the 5th, 10th, and 15th minutes, priority-based dynamic vertical scaling (Procedure~\ref{algo:autoscaling}) is executed. For DPM, the PS is first calculated for each server, after which scaling occurs. On the other hand for SPM, scaling based on the same initial PS is implemented. Updating priorities of 32 servers requires 30--40 seconds (s) and 15s for iPokeMon and FD, respectively. We infer from the figure that after the first scaling round, the SLO violation rates have decreased on average by 4\% and 3\% for iPokeMon and FD respectively. In subsequent scaling rounds, it is noted that the violation rates are further reduced. DPM on an average performs nearly 2\% better than SPM. The figure highlights that the sub-minute delay caused due to priority management and scaling does not necessarily affect the QoS of our use-cases.

\subsubsection{SLO Violation Rate}
Meeting SLOs is key to achieving a high overall QoS. Higher SLO violation rates at the Edge indicate the possibility of losing Edge customers. Consequently, this results in the loss of revenue for the Edge service provider. Therefore, the SLO violation rate is chosen to highlight the difference that DYVERSE makes to the QoS of Edge services.

From the empirical analysis, we observed that the average time to service an individual request for iPokeMon and FD is 78ms and 2.13s, respectively. {\color{black}The aim is to study the performance of DYVERSE when different tolerance for the SLO violation is accepted. This is useful for understanding in what situations, for example when the SLO is more stringent or lenient, DYVERSE can perform better.
%\todo{BV: Why is it important to know the impact on tolerance?}. 
Therefore, this study is pursued with SLOs that considers the average time taken to service a request and an upper tolerance of both 5\% and 10\% of the average time. For iPokeMon the SLOs are 78ms (0\% tolerance), 82ms (up to 5\% tolerance), 86ms (up to 10\% tolerance) and for FD the SLOs are 2.13s (0\% tolerance), 2.24s (up to 5\% tolerance) and 2.34s (up to 5\% tolerance).}

\begin{figure*}
\begin{center}
	\subfloat[SLO: 78 milliseconds]
	{\label{fig:figure3a}
	\includegraphics[width=0.3\textwidth]
	{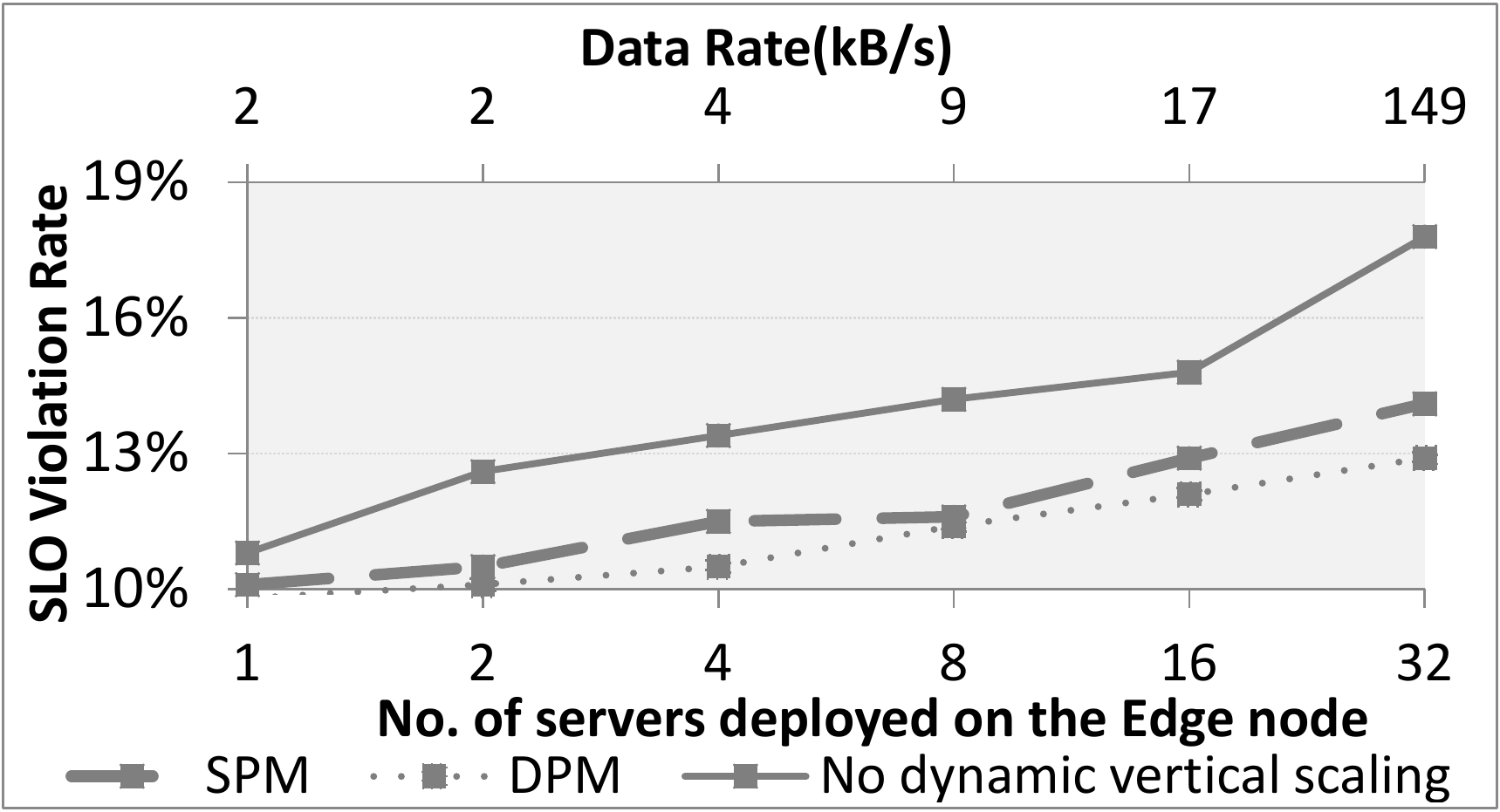}}
\hfill
	\subfloat[SLO: 82 milliseconds]
	{\label{fig:figure4a}
	\includegraphics[width=0.3\textwidth]
	{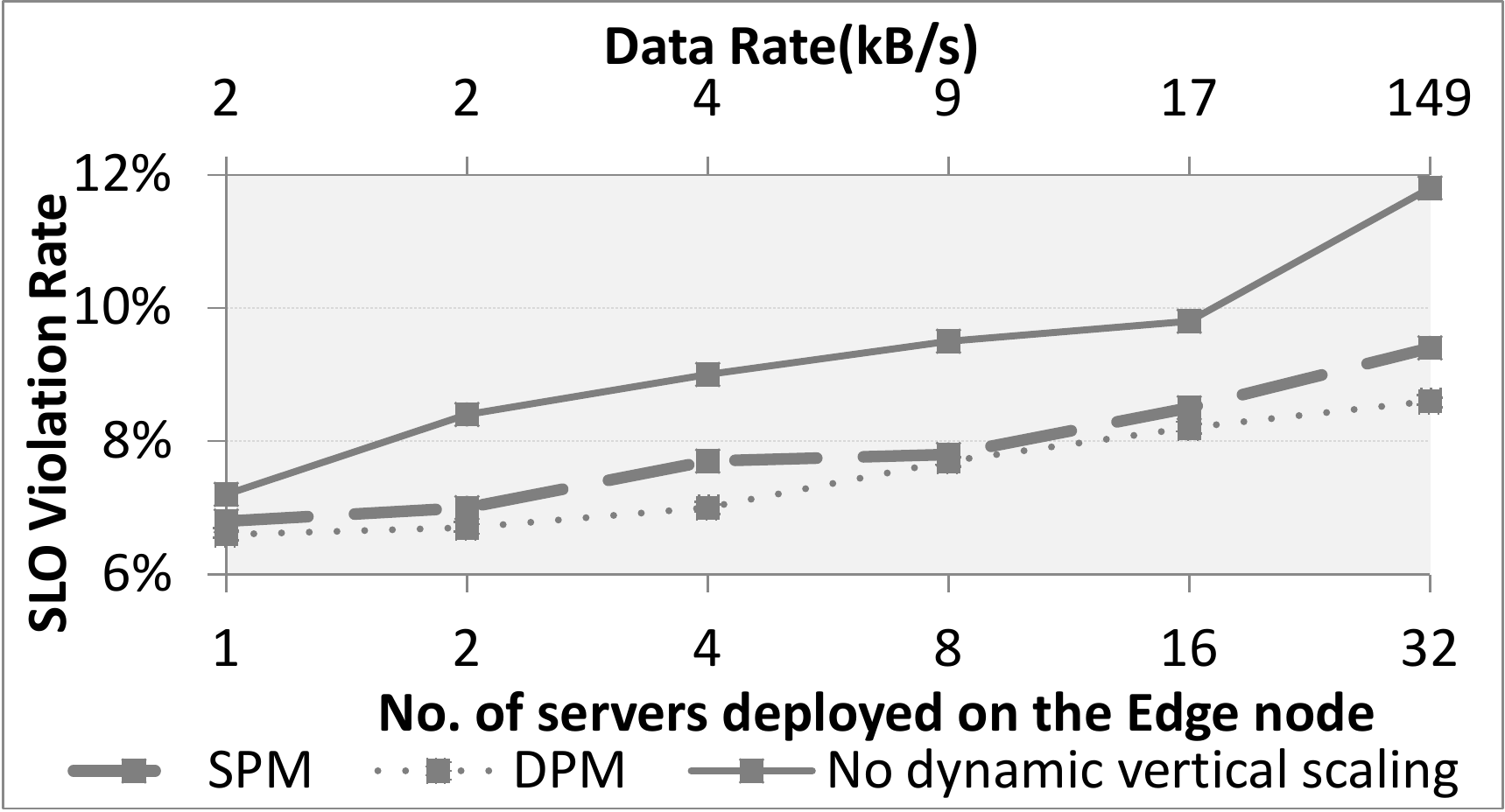}}
\hfill
	\subfloat[SLO: 86 milliseconds]
	{\label{fig:figure5a}
	\includegraphics[width=0.3\textwidth]
	{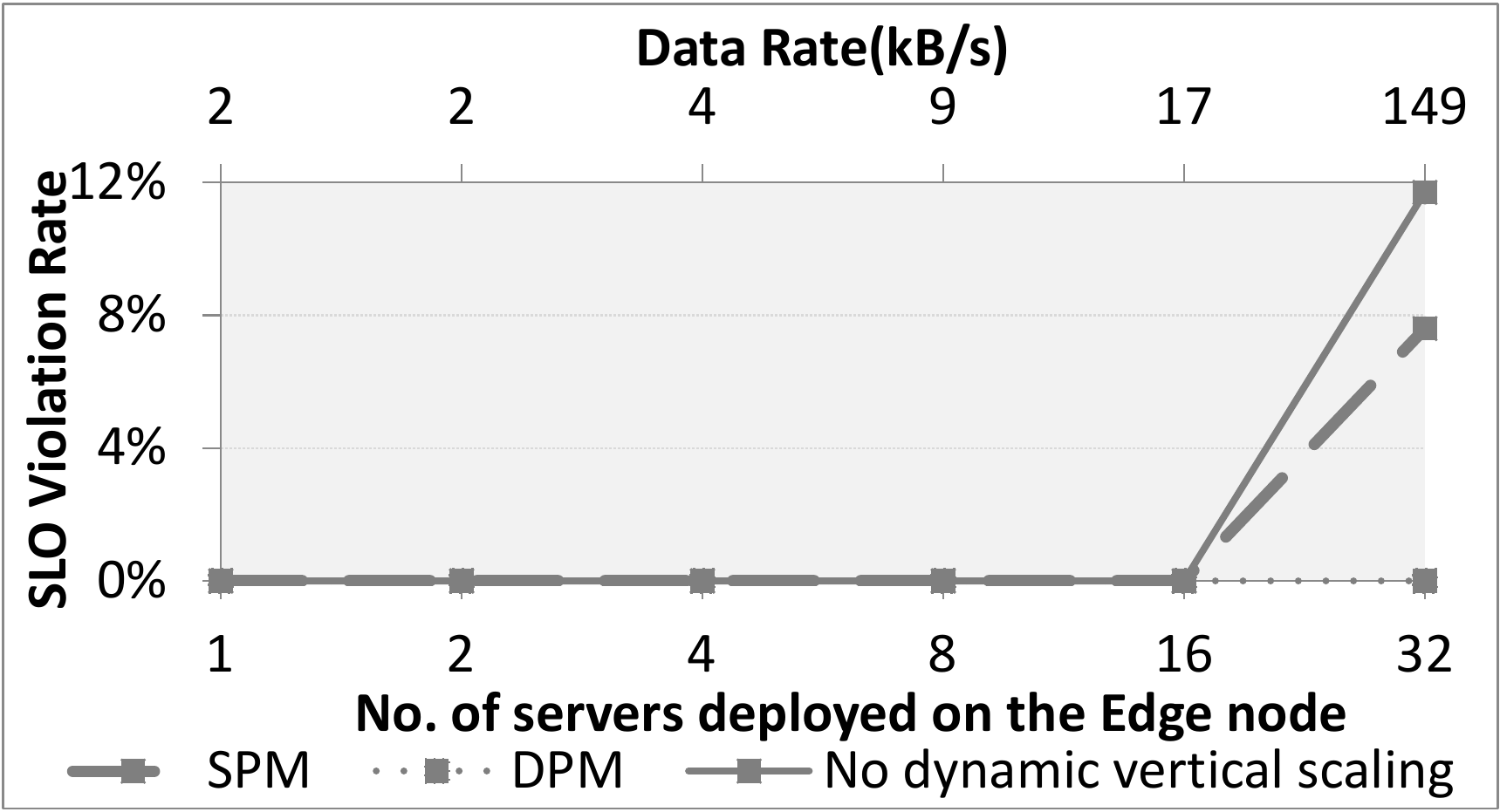}}
\end{center}
\caption{Violation rates of iPokeMon when applying varying SLOs.}
\label{fig:figure3}
\end{figure*}

\begin{figure*}
\begin{center}
	\subfloat[SLO: 2.13 seconds]
	{\label{fig:figure3b}
	\includegraphics[width=0.3\textwidth]
	{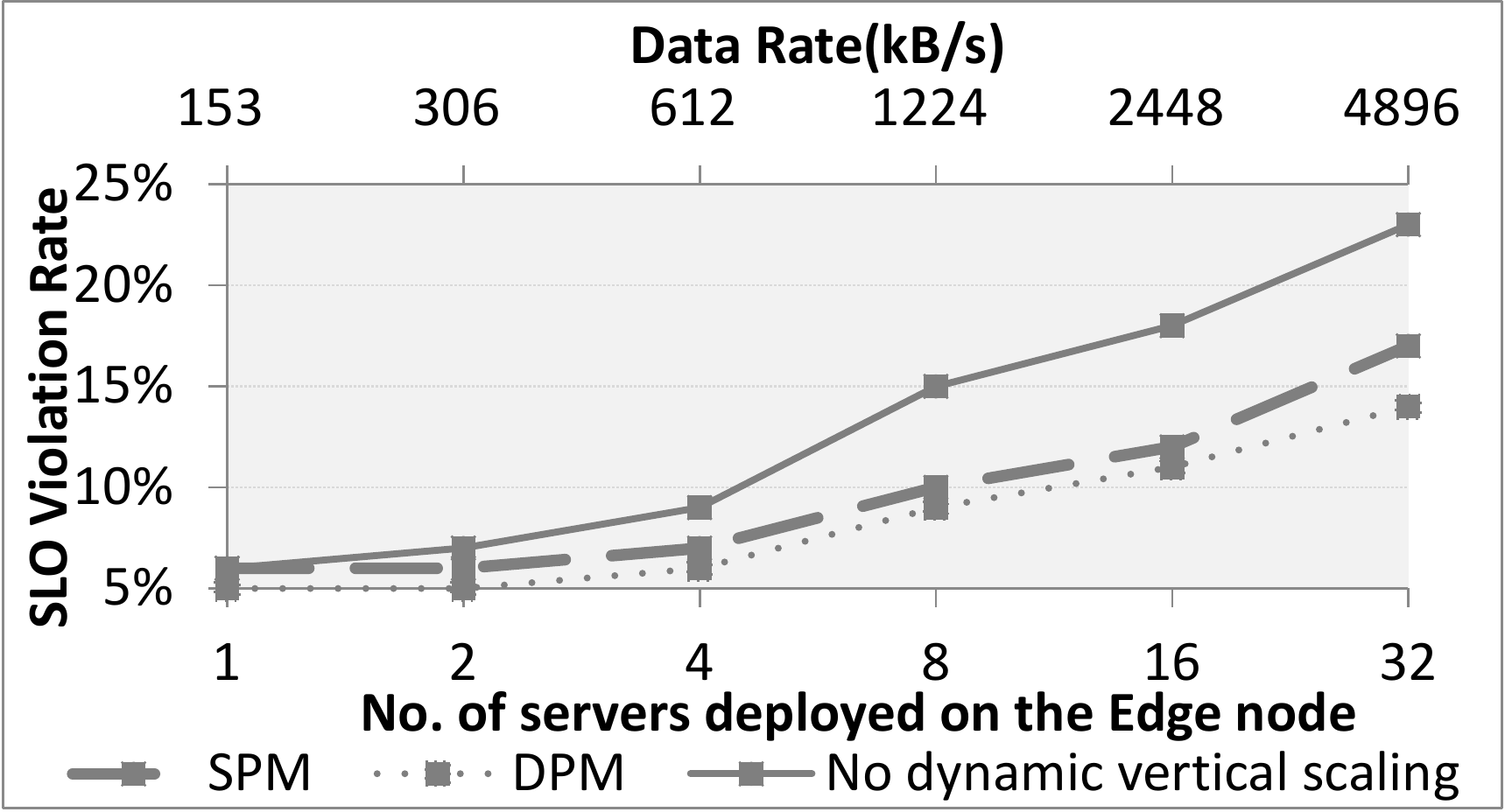}}
\hfill
	\subfloat[SLO: 2.24 seconds]
	{\label{fig:figure4b}
	\includegraphics[width=0.3\textwidth]
	{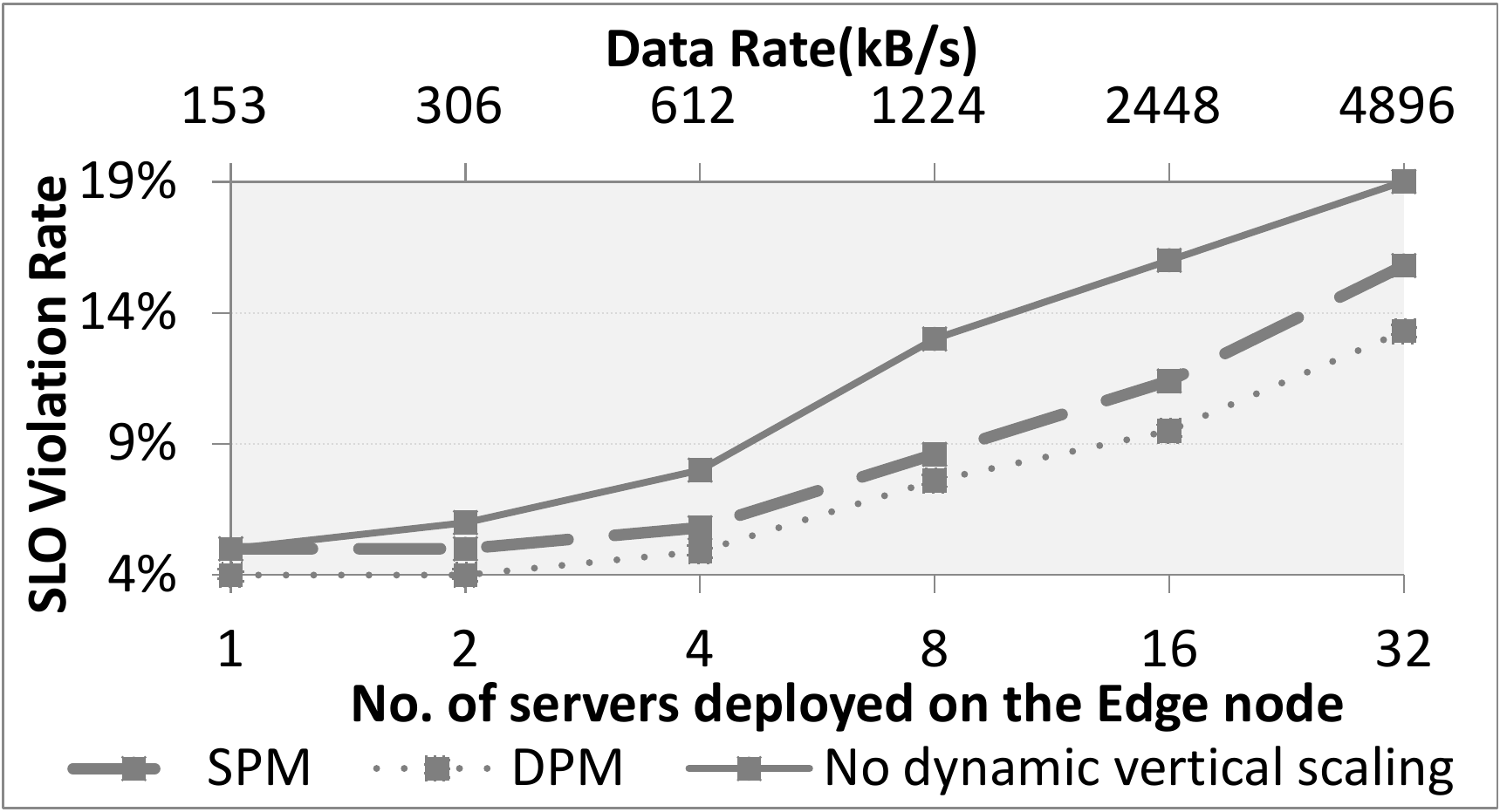}}
\hfill
	\subfloat[SLO: 2.34 seconds]
	{\label{fig:figure5b}
	\includegraphics[width=0.3\textwidth]
	{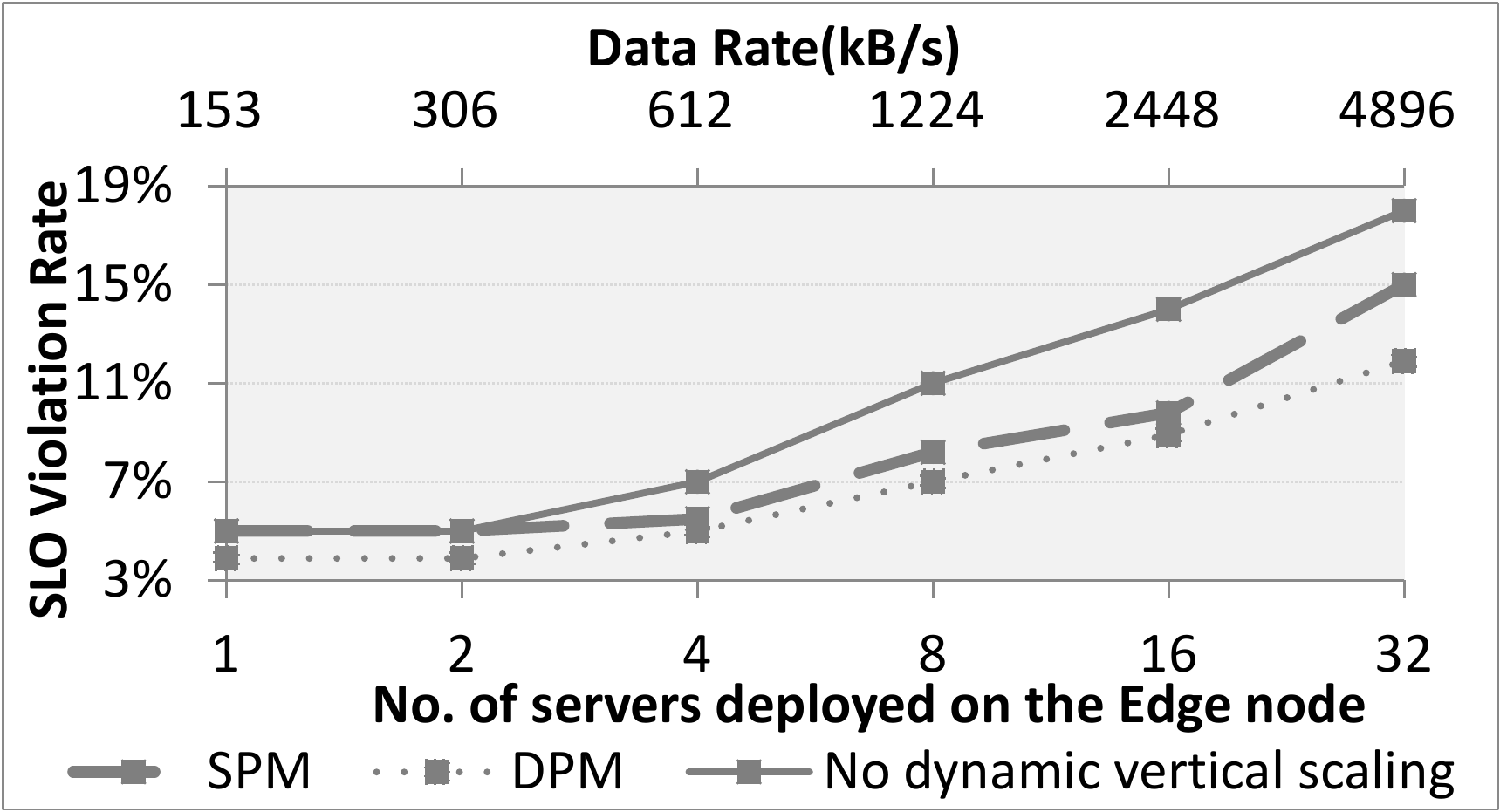}}
\end{center}
\caption{Violation rates of the face detection workload when applying varying SLOs.}
\label{fig:figure5}
\end{figure*}

The baseline we use for understanding the benefit of dynamic vertical scaling in static and dynamic priority management is by comparing them against a scenario in which there is no dynamic vertical scaling. {\color{black}This is because there are currently no alternative resource scaling solutions on a single Edge node.} We understand the benefit of using dynamic priority management by comparing it only against static priority management so that the benefit of the former is not magnified. 

{\color{black}Figure~\ref{fig:figure3} and Figure~\ref{fig:figure5} present the SLO violation rates of the two use cases when the SLO tolerates 0\% (Figures~\ref{fig:figure3a} and~\ref{fig:figure3b}), up to 5\% (Figures~\ref{fig:figure4a} and~\ref{fig:figure4b}) and up to 10\% (Figures~\ref{fig:figure5a} and~\ref{fig:figure5b}) violations of the average application latency.} The general trend is that for both workloads violation rates can be reduced by using the proposed scaling mechanism, either with SPM or DPM, when compared to a scenario without dynamic vertical scaling. When there is no dynamic vertical scaling, the containers hosting the Edge servers continue to retain the same amount of resources they were allocated before they started execution. No additional resources are dynamically allocated to containers. For example in Figure~\ref{fig:figure3a}, when the SLO is 78ms for 32 servers and 149~KBs of data is processed per second,
%and a total of 149 kilobytes~(kB) of data processed every second by the Edge node
nearly 18\% of the requests from users are violated. Scaling with SPM on an average reduces this violation by 4\%. A reduction of 6\% in SLO violation rate by applying SPM in FD is observed (Figure~\ref{fig:figure3b}) when 
%4,800~kB of data is processed every second by 
the SLO is 2.13s for 32 Edge servers and 4~MBs of data is processed per second.

The scaling mechanism takes the server's performance into account. During dynamic vertical scaling, the measured latency is compared with the user-defined SLO. If there were violations, then the container hosting the workload would be allocated additional resources and vice versa. Allocation to one container is done by requesting free resources from the Edge node or deallocating resources from other containers that have lower priorities. This optimises the use of Edge resources and can reduce the overall violation rates. {\color{black}This is noted for all SLOs for both workloads. For example, in Figure~\ref{fig:figure5b} for an SLO of 2.34s scaling with SPM reduces the violation rate for FD by 3\% over 32 servers (for iPokeMon with an SLO of 86ms in Figure~\ref{fig:figure5a} the violation rate is reduced by 4\% over 32 servers).}

However, the static priority of a server is set once before execution and does not change until it is terminated and needs to be executed again. This is a disadvantage since servers cannot change their priority during execution. Hence, we employ dynamic priorities that can be changed after deployment and during execution. Priorities can be implemented to take other factors, such as the number of users into account (in addition to factors listed in Table~\ref{table:staticPriority}, DPM considers factors presented in Table~\ref{table:dynamicPriority}). Resources can be scaled based on changing priorities. Dynamic priorities are better than static priorities since they further reduce the violation rates. When the violation threshold is stringent (SLO is 78ms for iPokeMon and 2.13s for FD), scaling with DPM reduces the violation by 1\% (Figure~\ref{fig:figure3a}) and 4\% (Figure~\ref{fig:figure3b}). Although we proposed three dynamic priorities (wDPS, cDPS, and sDPS) approaches, we have only considered sDPS for comparing the trends between no dynamic vertical scaling and scaling with SPM. This is because we observed that different approaches did not affect the overall violation rate. The effect of the three approaches is nonetheless considered in the context of latency.

When the violation threshold is lenient for iPokeMon (SLO is 86ms), a larger volume of requests can be fully serviced by the servers resulting in no violations for up to 32 Edge servers (Figure~\ref{fig:figure5a}). In these cases, requests that would result in a violation when the threshold is more stringent can now complete execution within the specified SLO. Subsequently, the benefits of DYVERSE becomes less important when the Edge node is less stressed. It is also noted that a scale-up operation presented in Procedure~\ref{algo:scaling} will not occur. However, containers may scale down during this time. {\color{black}With the same violation threshold, DPM for FD also results in a lower SLO violation rate, although the improvement is not as significant as in iPokeMon. This is because of larger data sizes - data transferred in FD is 30--150 times larger than data in iPokeMon. Images are frequently uploaded between the end device and the FD Edge servers, which results in longer processing times}.

\subsubsection{Latency of Workloads}

\begin{figure*}
\begin{center}
	\subfloat[SLO: 78 milliseconds]
	{\label{fig:figure6a}
	\includegraphics[width=0.3\textwidth]
	{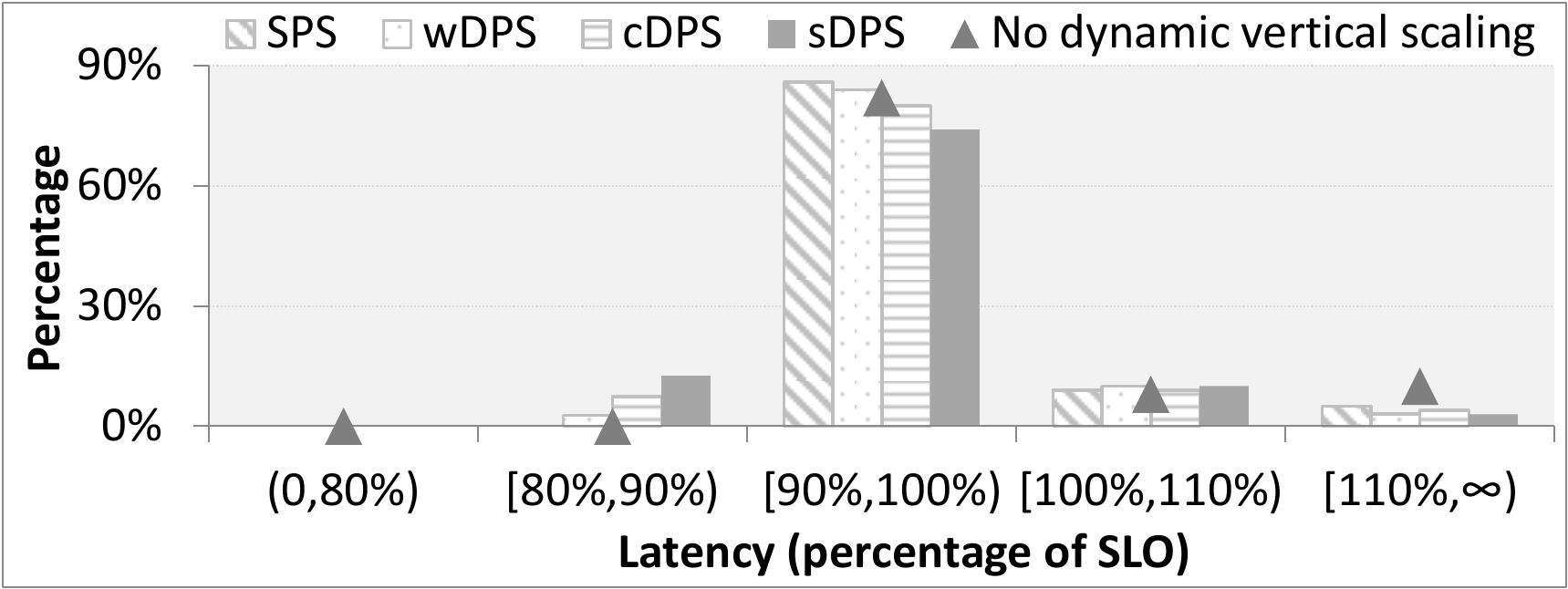}}
\hfill
	\subfloat[SLO: 82 milliseconds]
	{\label{fig:figure6b}
	\includegraphics[width=0.3\textwidth]
	{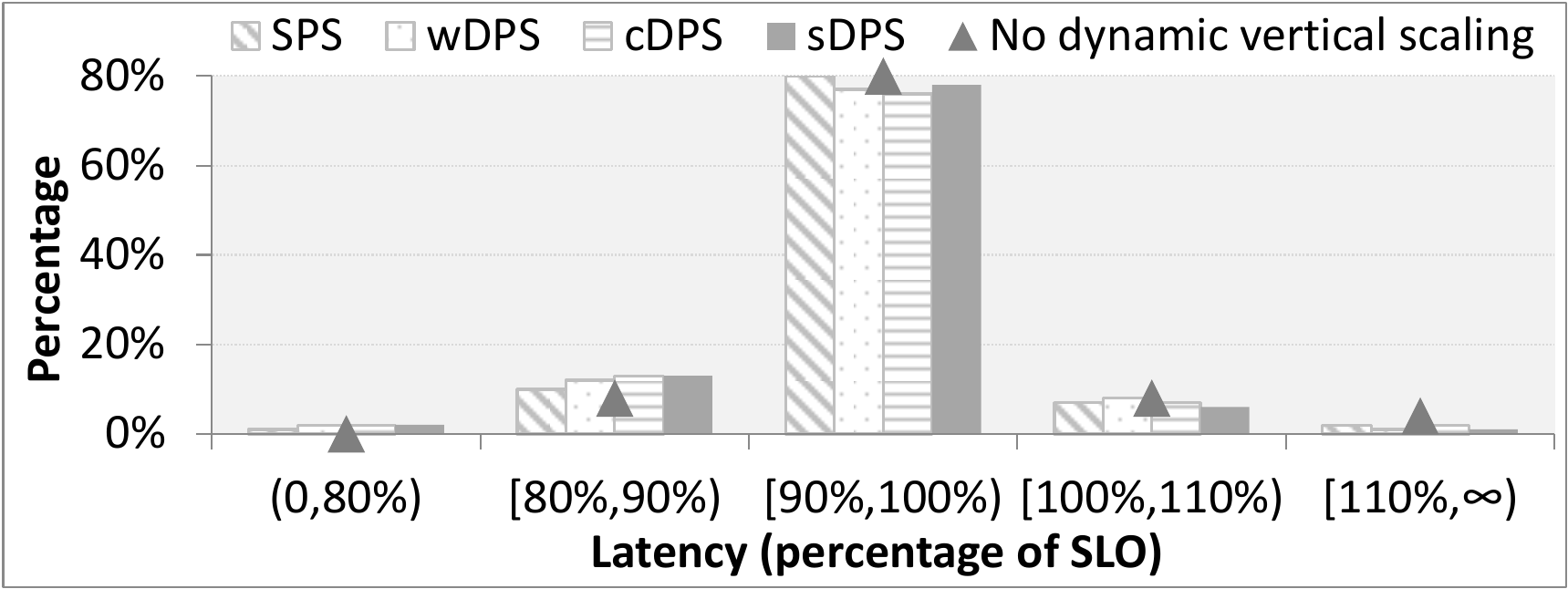}}
\hfill
	\subfloat[SLO: 86 milliseconds]
	{\label{fig:figure6c}
	\includegraphics[width=0.3\textwidth]
	{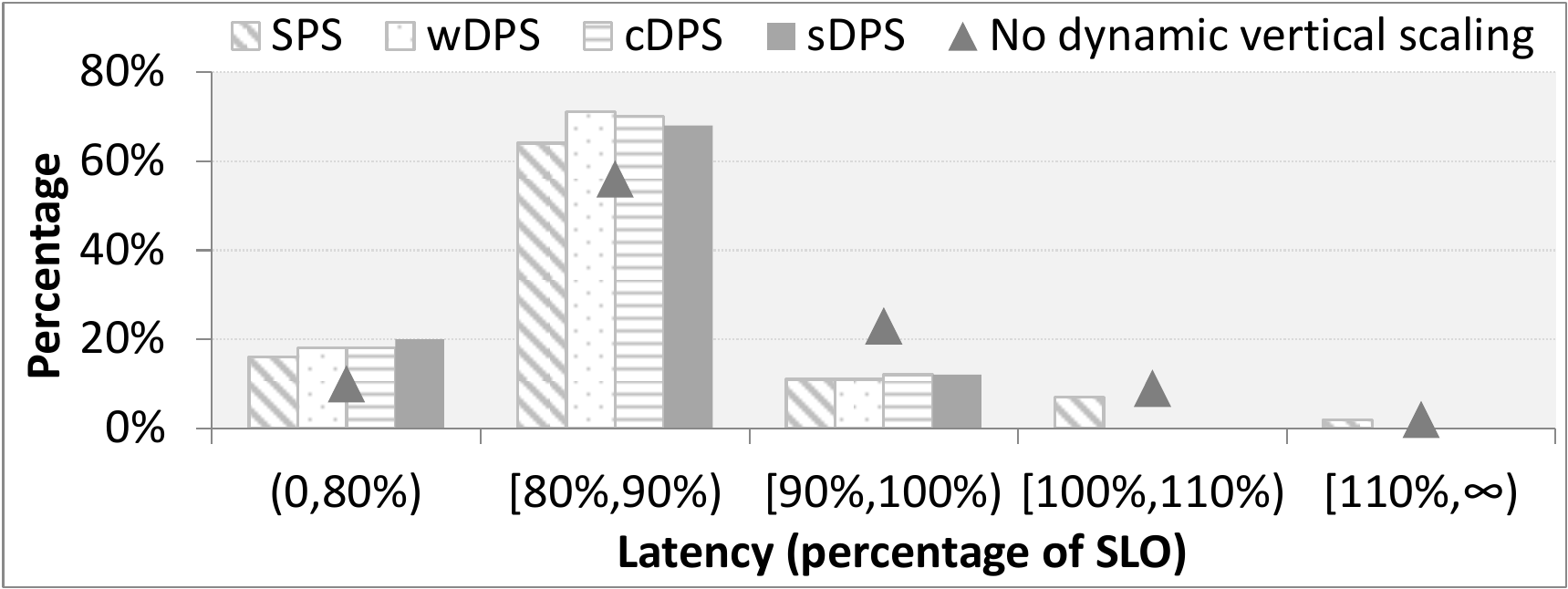}}
\end{center}
\caption{Latency of iPokeMon when 32 servers are deployed on the Edge node.}
\label{fig:figure6}
\end{figure*}

\begin{figure*}
\begin{center}
	\subfloat[SLO: 2.13 seconds]
	{\label{fig:figure7a}
	\includegraphics[width=0.3\textwidth]
	{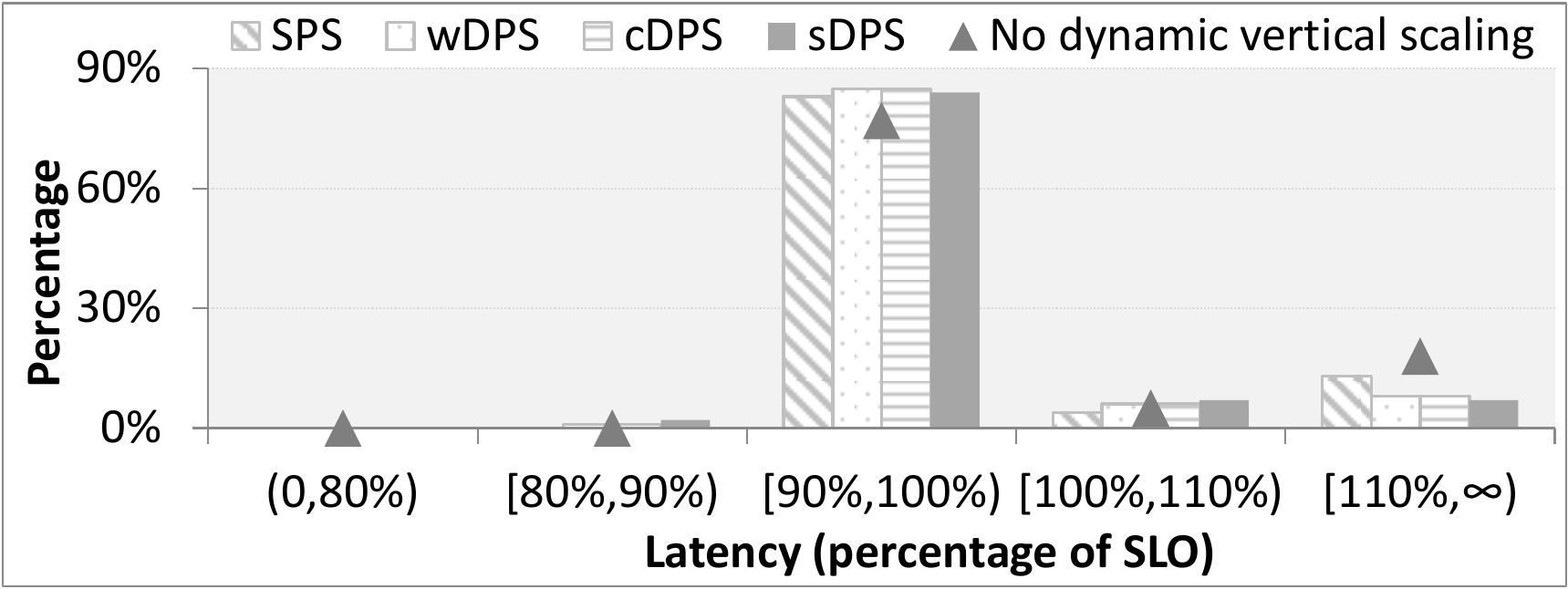}}
\hfill
	\subfloat[SLO: 2.24 seconds]
	{\label{fig:figure7b}
	\includegraphics[width=0.3\textwidth]
	{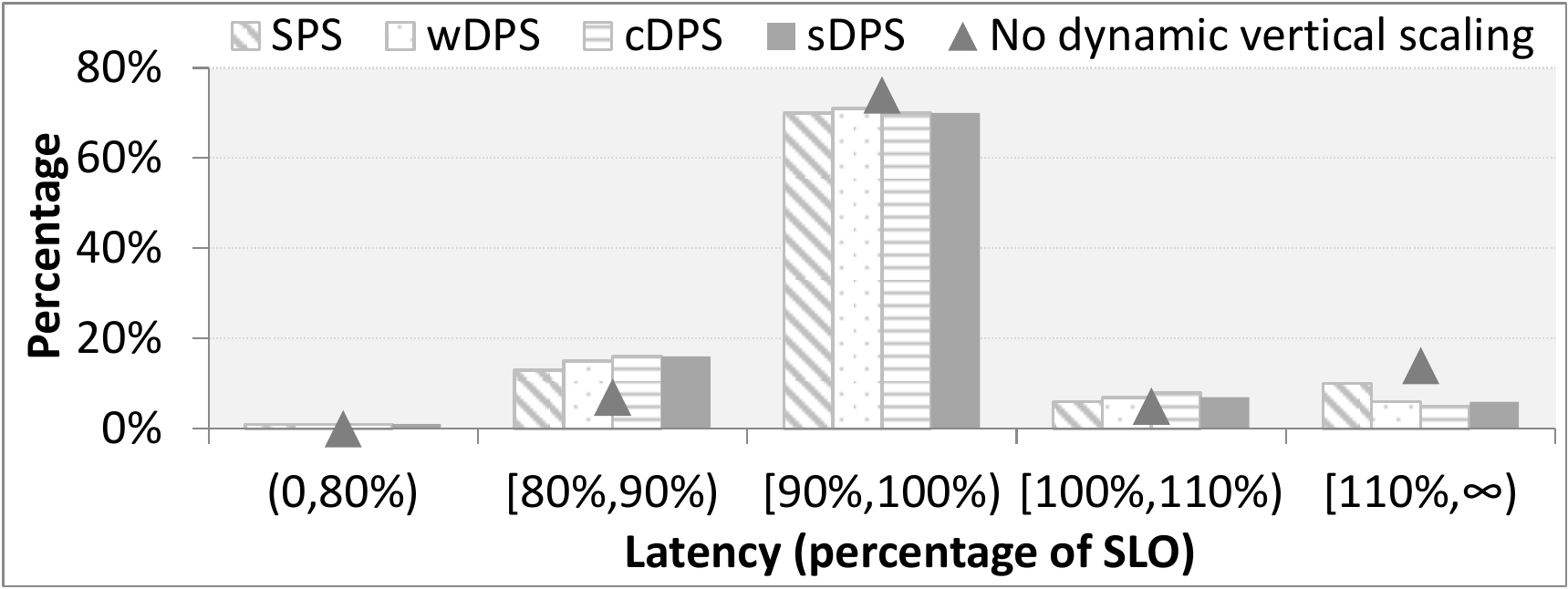}}
\hfill
	\subfloat[SLO: 2.34 seconds]
	{\label{fig:figure7c}
	\includegraphics[width=0.3\textwidth]
	{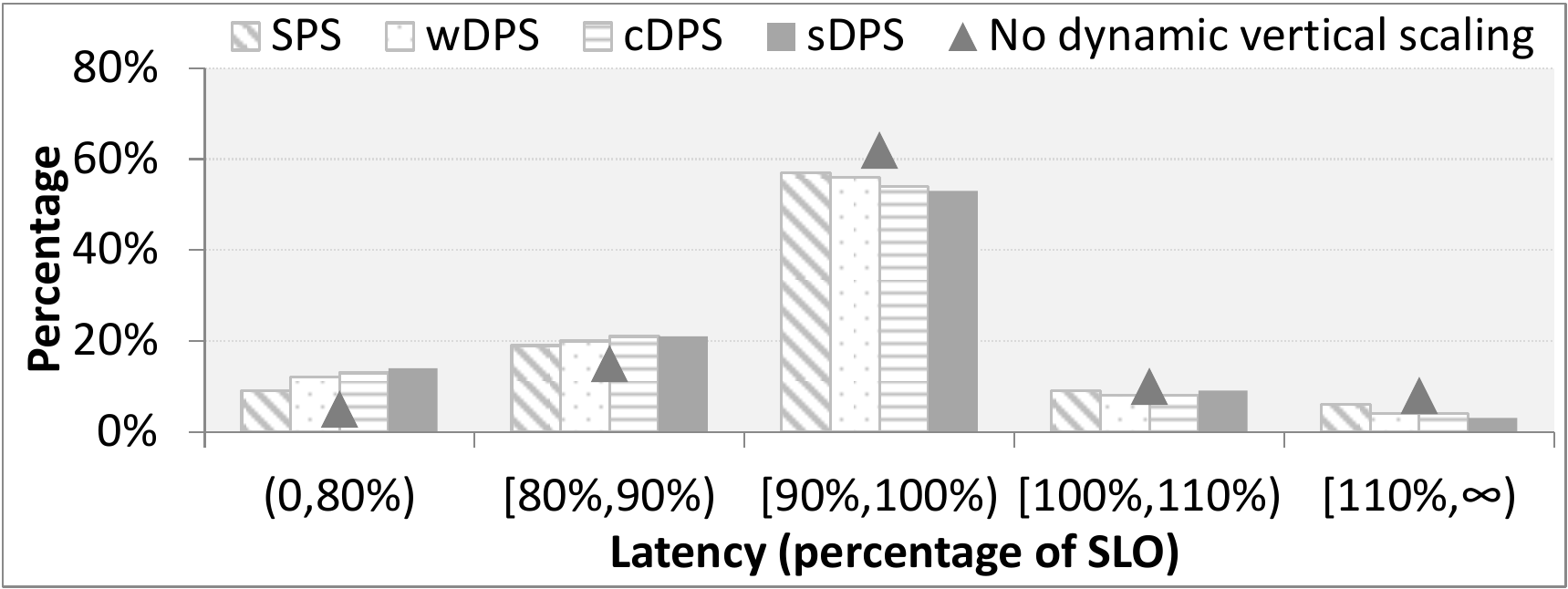}}
\end{center}
\caption{Latency of face detection when 32 servers are deployed on the Edge node.}
\label{fig:figure7}
\end{figure*}

The results shown above presented the violation rate when compared to the number of servers, but did not differentiate the three DPM approaches. This section aims to explore the impact of using different priority management approaches on latency. For iPokeMon, we define latency as the average round-trip time taken to service one request originating from the user by the Edge server. For FD, we define latency as the average single-trip time taken to detect faces by the Cloud server in one video frame originating from the camera. {\color{black}Figure~\ref{fig:figure6} and Figure~\ref{fig:figure7} show the distribution of application latency for the two use cases when the SLO tolerates 0\% (Figures~\ref{fig:figure6a} and~\ref{fig:figure7a}), 5\% (Figures~\ref{fig:figure6b} and~\ref{fig:figure7b}) and 10\% (Figures~\ref{fig:figure6c} and~\ref{fig:figure7c}) of the average application latency.} The distribution provides: (i) a time profile of requests violated, and (ii) the impact of DYVERSE on requests that are not violated. 

It is inferred that when the SLO violation threshold is 5\% and if no dynamic vertical scaling is used, then no requests are serviced in the lowest time band (from 0 to 80\% of the SLOs). However, the number of requests that are serviced in the lowest time band increases when employing static priority (the normal in the distribution shifts to the left). For example, for an SLO of 82ms for iPokeMon (Figure~\ref{fig:figure6b}), 4\% of requests are serviced within 66ms (the leftmost time band in the figure). This is due to the reallocation of resources from containers that meet SLOs to containers that violate SLOs in dynamic vertical scaling. {\color{black}Such a benefit is less obvious in FD, but a 1\% improvement is noted by applying dynamical vertical scaling for an SLO of 2.24s (Figure~\ref{fig:figure7b}).}

Dynamic priorities reduce the latency of both workloads when compared to static priority because they consider additional factors, such as the number of users 
%and  number of requests 
connected to each server. In most cases, sDPS performs better than cDPS and wDPS. For example, given an SLO of 86ms for iPokeMon (Figure~\ref{fig:figure6c}), 20\% of requests are serviced in the lowest time band using sDPS in contrast to 18\% for cDPS and wDPS. {\color{black}Similar results are observed in FD. For instance, given an SLO of 2.13s (Figure~\ref{fig:figure7a}), 2\% of video frames are processed by the face detector in the lowest time band using sDPS in contrast to 1\% for cDPS and wDPS.}
This is attributed to the penalty that is imposed on containers that scale. As presented previously, scaling incurs an overhead and therefore the Edge Manager aims to minimise the impact of scaling by reducing priorities as shown in Equation~\ref{eq:sdps}.

\subsubsection{Summary}
%\textit{\textbf{Summary}}: 
The results obtained from the experimental evaluation considered above are summarised as follows:

1) \textit{Overheads in using priorities and dynamic vertical scaling}: the overheads do not affect the execution of the servers and the servers continue to service user requests. In priority management, DPM takes a longer time if the monitoring of workload-related metrics is not incorporated in the Edge server (e.g. iPokeMon). Comparing with SPM, DPM has a 4\%--8\% higher overhead for dynamic vertical scaling. 
%When the number of servers on the Edge node increases, the average overheads of priority management and dynamic vertical scaling decrease since servers with lower priorities will be terminated.
%The priority-based dynamic scaling incurs a sub-minute delay, but it is observed that the QoS of applications are improved by a minimum 4\% reduction in SLO violation rates.

2) \textit{SLO violation rates for stringent thresholds (the average time to service one request)
}:
the SLO violation rate when there is no dynamic vertical scaling is observed to be nearly {\color{black} 18\% and 23\% for iPokeMon and FD respectively}. Scaling with SPM improves the violation rate by {\color{black}4\% and 6\% for iPokeMon and FD respectively} when compared to no dynamic vertical scaling. 
Scaling with DPM improves the violation rate of iPokeMon by 1\% 
%and 8\% for stringent and lenient SLOs, respectively, 
when compared to SPM. {\color{black}This is confirmed by the 3\% reduction of SLO violation rate of the face detection workload when comparing DPM and SPM.} 

3) \textit{SLO violation rates for lenient thresholds (10\% more of the average time taken to service one request)}: the SLO violation rate when there is no dynamic vertical scaling is observed to be nearly {\color{black}12\% and 18\% for iPokeMon and FD respectively}. Scaling with SPM improves the violation rate by {\color{black}3\% for FD and 4\% for iPokeMon} when compared to no dynamic vertical scaling. Scaling with DPM ensures no requests are violated for iPokeMon and improves the violation rate by 3\% for FD when compared to SPM.

4) \textit{Latency for stringent thresholds
%(the average time to service one request)
}:
without dynamic vertical scaling, no requests are serviced within 66ms for iPokeMon {\color{black}and within 2.13s for FD}. Using sDPS with iPokeMon, nearly 13\% of requests are serviced in 66ms--74ms, which is higher than cDPS and wDPS by 5\% and 10\% respectively. {\color{black}Using sDPS with FD offers a small advantage; 1\% more requests are processed in 1.79--2.02s}. DPM performs better than SPM. Using sDPS, 20\% of requests of iPokeMon are serviced within 72ms, which is higher than SPS, cDPS, and wDPS by 4\%, 2\% and 2\% respectively. {\color{black}Similarly for FD, 2\% of video frames are processed in 1.7s--1.92s, which is higher than SPS, cDPS and wDPS by 2\%, 1\% and 1\% respectively}.

5) \textit{Latency for lenient thresholds
%(10\% more of the average time taken to service one request)
}: Fewer requests are serviced within 69ms for iPokeMon {\color{black}and 1.87s for FD} with no dynamic vertical scaling compared to using it. Using sDPS with iPokeMon, 20\% of requests are serviced in the lowest time band, which is higher than cDPS and wDPS by 2\%. {\color{black}Using sDPS with FD, 14\% of requests are serviced in the lowest time band, which is higher than cDPS and wDPS by 1\% and 2\% respectively.} DPM performs better than SPM.

%% file: sections/relatedwork.tex
%\begin{figure}
%\begin{center}
%	\includegraphics[width=\textwidth]
%	{images/RelatedWork.png}
%\end{center}
%\caption{A classification of existing research on resource scaling.}
%\label{fig:relatedwork}
%\end{figure}

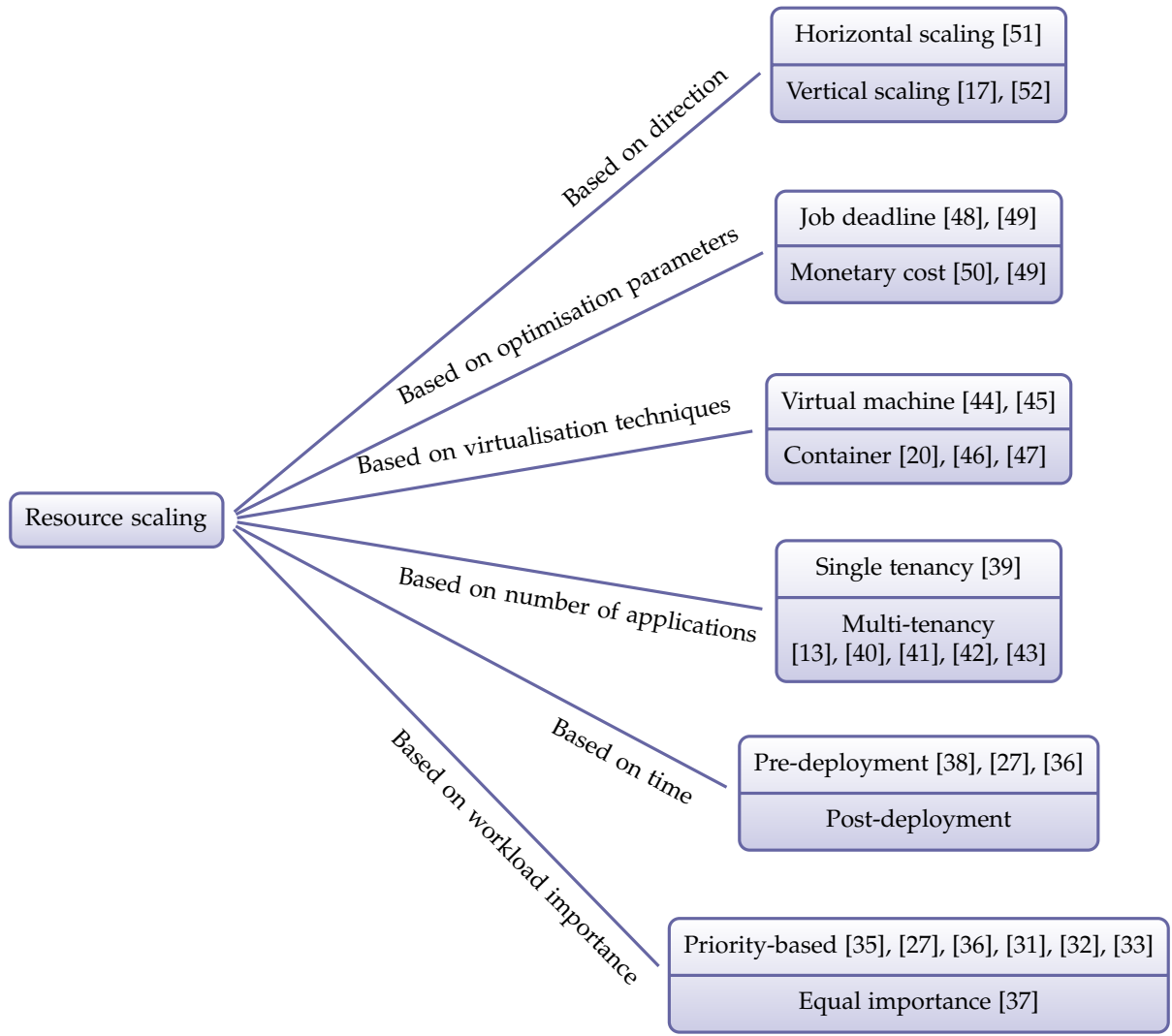
\begin{figure*}
\centering
\begin{tikzpicture}[
    grow=right,
    level 1/.style={sibling distance=2.5cm,level distance=11cm},
    level 2/.style={sibling distance=3.5cm, level distance=6.7cm},
    edge from parent/.style={very thick,draw=blue!40!black!60,
        shorten >=5pt, shorten <=5pt},
    edge from parent path={(\tikzparentnode.east) -- (\tikzchildnode.west)},
    kant/.style={text width=10cm, text centered, sloped},
    every node/.style={text ragged, inner sep=2mm},
    punkt/.style={rectangle, rounded corners, shade, top color=white,
    bottom color=blue!50!black!20, draw=blue!40!black!60, very
    thick }
    ]

\node[punkt, text width=7.5em] {Resource scaling}
    child {
        node[punkt] [rectangle split, rectangle split parts=2,
         text ragged] {
            Priority-based~\cite{you2017energy, yu2016joint, tortonesi2019taming, chen2013user, he2012elastic, pastorelli2015hfsp}
                  \nodepart{second}
            Equal importance~\cite{buyya2002economic}
        }
        edge from parent
            node[kant, below, pos=.7] {Based on workload importance}
    }
    child {
        node[punkt] [rectangle split, rectangle split parts=2,
         text ragged] {
            Pre-deployment~\cite{brogi2017qos, yu2016joint, tortonesi2019taming}
                  \nodepart{second}
            Post-deployment
        }
        edge from parent
            node[kant, below, pos=.8] {Based on time}
    }
    child {
        node[punkt] [rectangle split, rectangle split parts=2,
         text ragged, align=center] {
            Single tenancy~\cite{chieu2009dynamic}
                  \nodepart{second}
            Multi-tenancy\\\cite{liu2014reciprocal, shen2011cloudscale, paulraj2018resource,  yang2015multi, lyu2017multiuser}
        }
        edge from parent
            node[kant, below, pos=.65] {Based on number of applications}
    }
    child {
        node[punkt] [rectangle split, rectangle split parts=2,
         text ragged, align=center] {
            Virtual machine~\cite{yazdanov2014lightweight, mao2011auto}
                  \nodepart{second}
            Container~\cite{Thoth2017, zhang2019quantifying, abdul2019performance}
        }
        edge from parent
            node[kant, above, pos=.6] {Based on virtualisation techniques}
    }
    child {
        node[punkt] [rectangle split, rectangle split parts=2,
         text ragged] {
            Job deadline~\cite{marshall2010elastic, jiang2013optimal}
                  \nodepart{second}
            Monetary cost~\cite{long-1, jiang2013optimal}
        }
        edge from parent
            node[kant, above, pos=.65] {Based on optimisation parameters}
    }
    child {
        node[punkt] [rectangle split, rectangle split parts=2,
         text ragged] {
            Horizontal scaling~\cite{horizontalscaling-1}
                  \nodepart{second}
            Vertical scaling~\cite{verticalscaling-1, li2016load}
        }
        edge from parent
            node[kant, above, pos=.8] {Based on direction}
    }
    ;
\end{tikzpicture}
\caption{\color{black}A classification of existing research on resource scaling.}
\label{fig:relatedwork}
\end{figure*}

Resource scaling is well studied in distributed systems and more recently on the Cloud. {\color{black}A classification of existing research is shown in Figure~\ref{fig:relatedwork}. In this article, six classifications are highlighted based on (i) the direction of scaling, (ii) the parameters that are optimised in the problem space, (iii) the virtualisation techniques that the resource scaling solutions target at, (iv) the number of applications the scaling algorithms consider, (v) the time when the resource scaling is performed, and (vi) the importance of workload being considered.

\textit{Based on the direction of resource scaling}, computing resources for a workload can be allocated or deallocated from a cluster either horizontally~\cite{horizontalscaling-1} or vertically~\cite{verticalscaling-1}. Horizontal scaling, also referred to as scale-out/scale-in, is an action to add/remove infrastructure capacity in pre-packaged blocks of resources. In addition to horizontal scaling, vertical scaling, also referred to as scale-up/scale-down, is an action to add/remove resources to an existing system~\cite{verticalscaling-1}. }

{\color{black}
The focus of this paper is vertical scaling on the Edge. The vertical scaling approaches used on the Cloud are not suited for the Edge due to three reasons. Firstly, monitoring techniques employed for vertical scaling are heavyweight (i.e., time consuming and require significant resources)~\cite{spinner2014runtime}, which cannot be directly applied to resource-limited Edge environments (a large cluster of resources may not be available at the edge of the network). Secondly, generating optimal scheduling plans for vertical scaling as employed in the Cloud~\cite{dutta2012smartscale} is prohibitive on the Edge. Integer linear programming or constraint programming that is usually used to generate optimal solutions are computationally intensive. Thirdly, vertical scaling on the Cloud is underpinned by workload prediction models~\cite{li2016load}, which are less suitable for use on the Edge. Edge services are likely to be short-running workloads in contrast to the Cloud and only limited data may be available for training machine learning models. Additionally, training models on the Edge may not be feasible. Therefore, in this paper, we propose a low overhead vertical scaling technique on an Edge node.}

{\color{black}
\textit{Based on the parameters that are optimised} in the problem space, two important parameters are frequently considered, namely job deadline and monetary cost. Workload deadline has been employed as an important parameter to optimise during resource scaling~\cite{marshall2010elastic, jiang2013optimal}, as it directly represents the application performance. Moreover, cost budget also plays an important role~\cite{long-1, jiang2013optimal} because in many cases the more resources being utilised, the higher the monetary cost it entails.

\textit{Based on the virtualisation techniques being targeted}, existing research has proposed resource scaling algorithms for both virtual machines~\cite{yazdanov2014lightweight} and containers~\cite{Thoth2017, zhang2019quantifying}.} Given different types of VMs with variable CPU and memory resources, sometimes workloads are simply migrated onto another VM with resources more suitable to meet the optimising criteria (instead of scaling the resource on the original VM)~\cite{mao2011auto}.

{\color{black}\textit{Based on the number of applications being considered} in the computing environment, resource scaling solutions have been designed for single tenancy~\cite{chieu2009dynamic} and more often multi-tenancy~\cite{shen2011cloudscale}.} Multi-tenancy refers to the co-location of different workloads on the same computing resource. In a multi-tenant Cloud environment, migration is employed to resolve resource contention~\cite{shen2011cloudscale, paulraj2018resource}. Inter-tenant resource trading and intra-tenant weight adjustments are employed on the Cloud to ensure fairness when scaling~\cite{liu2014reciprocal}.

Techniques employed on the Edge to deal with multi-tenancy need to be again lightweight and cannot incorporate complex models for workload predictions~\cite{hong2018resource}. 
Complex schemes, such as convex optimisation, game theory or dynamic programming have been explored for Edge environments~\cite{mao2017survey}. Heuristic-based mobile application partitioning and offloading algorithms have been designed to maximise the application's performance for multiple users~\cite{yang2015multi} and the system utility~\cite{lyu2017multiuser}. These algorithms are scalable but still require $O(K^2), O(K^3)$ time respectively, to service $K$ mobile users. This is impractical on resource-constrained Edge nodes that use small processors since they cannot acquire hardware resources easily. Similarly, game theory based schemes have been designed for Mobile-Edge Cloud computing~\cite{chen2016efficient}. This approach is tested on a cloudlet, which naturally assumes more resources than the extreme edge of the network (cloudlets have dedicated computing servers/clusters). 

\textit{Based on the time when resource scaling actions are performed}, the adjustment of resource allocation could happen either before or after the deployment of applications.
{\color{black}Research on multi-tenancy at the Edge usually focuses on pre-deployment. Once a workload is deployed, the resources allocated to it do not change. Similarly, the QoS of applications is improved without considering multi-tenancy after deployment~\cite{brogi2017qos}. To the best of our knowledge, research on post-deployment resource scaling in Edge computing is sparse at best. However, multi-tenancy significantly impacts the SLO violation rate after deployment. In contrast to existing research that is reported in the literature on multi-tenancy in Edge computing, this research accounts for post-deployment conditions of workloads for scaling resources in a multi-tenant environment. This paper demonstrates that dynamic vertical scaling can improve the overall QoS of the application during its lifetime on the Edge after it is deployed.} 

\textit{Based on the importance of workloads}, some resource scaling algorithms are priority-based whereas others treat the application equally important. Community models are used to develop a shared resource pool in Clouds~\cite{buyya2002economic}. 
The priority of Edge workloads is different because relationships between (i) the Cloud and the Edge, and (ii) the user and the Edge need to be accounted for. There is research that (i) adopts a threshold-based priority function to decide on partial or complete offloading of workloads~\cite{you2017energy}, {\color{black}and (ii) adjusts the task execution order of workloads to satisfy different objectives in the pre-deployment phase~\cite{yu2016joint, tortonesi2019taming}. These show the impact of priorities on offloading, but do not demonstrate the influence of priorities on post-deployment resource management.} In addition, existing resource management solutions that consider workload priority only takes a single factor that impacts the priority, for example, the purchased priority~\cite{chen2013user}, the sequence of incoming workloads~\cite{he2012elastic} or the potential resource starvation~\cite{pastorelli2015hfsp}. To advance the state-of-the-art, the research reported in this paper considers a collection of static factors that are specific to the Edge, as well as dynamic factors that vary during the execution of workloads. These factors all together contribute to the workload priorities for efficiently managing resources in a multi-tenant environment.

%% file: sections/conclusions.tex
Distributed applications will leverage the edge of the network to improve their overall QoS for which the challenge of multi-tenancy in resource-constrained environments need to be addressed.
Vertical scaling of resources is required to achieve multi-tenancy.
However, existing mechanisms require significant monitoring and are computationally intensive since they were designed for the Cloud. They are not suitable for resource-limited Edge. 

The research in this paper addresses the above problem by developing DYVERSE, a \textbf{DY}namic \textbf{VER}tical \textbf{S}caling mechanism on \textbf{E}dge environments to facilitate multi-tenancy. The aim is to maximise the QoS of workloads executing in a multi-tenant Edge node without violating user-defined Service Level Objectives (SLOs). The proposed scaling mechanism generates post-deployment plans for workloads during execution so that SLOs are not violated. The mechanism is underpinned by one static and three dynamic priority management approaches. Three dynamic priorities that are workload-aware, community-aware and system-aware are proposed. %We hypothesised that dynamic vertical scaling along with priority management improves the QoS of multi-tenant workloads and reduces SLO violation rates. 

The feasibility of DYVERSE is validated using two workloads, namely a location-aware online game and a real-time face detection application in a Cloud-Edge environment. The priority-based dynamic vertical scaling has less than a sub-second overhead per server when 32 Edge servers are deployed on a single Edge node. Scaling with static priorities reduced SLO violation rates of user requests by up to 4\% and 12\% for the online game respectively and in both cases 6\% for the face detection workload, when compared to no dynamic vertical scaling. For both workloads, system-aware dynamic vertical scaling effectively reduced the latency of non-violated requests, when compared to other methods.
%All the requests were executed without violations using scaling with dynamic priorities when the SLO is lenient. When compared to an approach that does not employ dynamic vertical scaling, the system-aware dynamic priority approach improved the application round-trip latency for requests that are not violated. 
%The key result is that the initial hypothesis was confirmed. 

\subsection{Limitations and Future Work}
%\textbf{\textit{Limitations and Future Work}}:
Although system-related parameters that affect SLO violation rates and latency are considered, network-based parameters, such as bandwidth and how it affects the QoS were not considered. This is a part of our ongoing research and will be reported elsewhere.   

%The priority of each server is currently derived from a linear combination assuming that all factors have equal weights. This is suitable in the context of a proof-of-concept multi-tenant Edge environment. Varied weights for the factors that affect priority will provide a more fine-grained approach to control the multi-tenant environment. In the future, we aim to integrate weights to define priorities and understand its implications on workloads. 
{\color{black}An extensive study of the direct impact of different configurations of the factor weights on the system performance is not included in this paper. The current experimental results demonstrate a proof-of-concept of the feasibility of multi-tenant Edge environments. In the future, we aim to integrate the tuning of weights to optimise system performance.}

%Additional variables that can be obtained from monitoring hardware resources on the Edge node can be included. However, this will result in a more heavyweight approach for priority management and dynamic vertical scaling. Efforts will be made to explore how additional variables can be included in our model to optimise the QoS problem as well as the trade-off between QoS and the overheads. 
The dynamic vertical scaling mechanism proposed in this paper considers only QoS metrics of an Edge server, namely workload latency and SLO violation rate, to invoke resource scaling. Utilisation metrics of containers, such as memory and swap space used for hosting Edge servers, could be further considered.

%We also aim to develop a broader range of priority management approaches. For example, in the community-aware DPS approach, an alternate economic model that relies on barter so that tenants exchange resources without donating them to a shared pool will be investigated.